%% file: 00-main.tex
\documentclass[sigconf, 10pt]{acmart}
\AtBeginDocument{%
  \providecommand\BibTeX{{%
    \normalfont B\kern-0.5em{\scshape i\kern-0.25em b}\kern-0.8em\TeX}}}

\acmYear{2024}\copyrightyear{2024}
\setcopyright{acmlicensed}
\acmConference[ACM MobiCom '24]{International Conference On Mobile Computing And Networking}{September 30--October 4, 2024}{Washington D.C., DC, USA}
\acmBooktitle{International Conference On Mobile Computing And Networking (ACM MobiCom '24), September 30--October 4, 2024, Washington D.C., DC, USA}
\acmDOI{10.1145/3636534.3649352}
\acmISBN{979-8-4007-0489-5/24/09}

\usepackage{xspace}
\usepackage{enumitem}
\usepackage{graphicx}
\usepackage{subcaption}
\usepackage{multirow}
\usepackage{bbding}
\usepackage{adjustbox}
\usepackage{hyperref}
\usepackage{algpseudocode}
\usepackage{algorithmicx}
\usepackage{algorithm}
\usepackage{amsmath}
\usepackage{multirow}  
\usepackage{enumerate}
\usepackage{pifont}
\usepackage[normalem]{ulem}
\usepackage{diagbox}

\newcommand{\sysname}{\textit{Soar}\xspace}
\newcommand{\myparagraph}[1]{\vspace{0mm} \noindent \textbf{\textrm{#1}}}

\newcommand{\blue}[1]{\textcolor{black}{#1}}

\newcommand{\new}[1]{\textcolor{black}{#1}}

\begin{document}

\title{\sysname: Design and Deployment of A Smart Roadside Infrastructure System for Autonomous Driving}
\renewcommand{\shorttitle}{\sysname: Design and Deployment of A SRI System for Autonomous Driving}

\author{Shuyao Shi\texorpdfstring{$^{*}$}{}, Neiwen Ling\texorpdfstring{$^{*}$}{}, Zhehao Jiang\texorpdfstring{$^{*}$}{}, Xuan Huang\texorpdfstring{$^{*}$}{}, Yuze He, Xiaoguang Zhao, Bufang Yang, Chen Bian, Jingfei Xia, Zhenyu Yan, Raymond W. Yeung, and Guoliang Xing\texorpdfstring{$^{\dagger}$}{}}
\affiliation{\institution{{}The Chinese University of Hong Kong, Hong Kong SAR, China} \country{}}
\thanks{*Co-primary authors.}
\thanks{$^{\dagger}$Corresponding author.}
\renewcommand{\shortauthors}{Shi, Ling, Jiang, and Huang, et al.}

\begin{abstract}
Recently, \textit{smart roadside infrastructure (SRI)} has demonstrated the potential of achieving fully autonomous driving systems. To explore the potential of infrastructure-assisted autonomous driving, this paper presents the design and deployment of \sysname, the first end-to-end SRI system specifically designed to support autonomous driving systems. \sysname consists of both software and hardware components carefully designed to overcome various system and physical challenges. \sysname can leverage the existing operational infrastructure like street lampposts for a lower barrier of adoption. \sysname adopts a new communication architecture that comprises a bi-directional multi-hop I2I network and a downlink I2V broadcast service, which are designed based on off-the-shelf 802.11ac interfaces in an integrated manner. \sysname also features a hierarchical DL task management framework to achieve desirable load balancing among nodes and enable them to collaborate efficiently to run multiple data-intensive autonomous driving applications. We deployed a total of 18 \sysname nodes on existing lampposts on campus, which have been operational for over two years. Our real-world evaluation shows that \sysname can support a diverse set of autonomous driving applications and achieve desirable real-time performance and high communication reliability. Our findings and experiences in this work offer key insights into the development and deployment of next-generation smart roadside infrastructure and autonomous driving systems. 
\end{abstract}




\maketitle

\input{01-Intro.tex}

\input{02-Related.tex}
\input{03-Overview.tex}

\input{04-Communication}
\input{05-Computing.tex}

\input{06-Deployment.tex}

\section{Evaluation}
\label{sec:evaluation}
\input{07-1-Exp-Setup}
\input{07-2-Exp-Overall}

\input{07-3-Exp-comm}
\input{07-4-Exp-sys-eva}

\input{08-Discussion}
\input{09-Conclusion}

%% Acknowledgments
\begin{acks}
This work is supported in part by the Research Grants Council (RGC) of Hong Kong under General Research Fund under Grant 14222222, and the Centre for Perceptual and Interactive Intelligence (CPII) under Grant EW01610 (RP4-3).
\end{acks}

\bibliographystyle{ACM-Reference-Format}
\bibliography{reference}

\end{document}

%% file: 01-Intro.tex
\section{Introduction}
\label{sec:intro}
Autonomous driving is envisioned to revolutionize our transportation system. However, current pilot commercial deployments have posed major concerns regarding the safety of existing autonomous driving systems. Recent years have witnessed the emergence of a new paradigm that leverages \textit{smart roadside infrastructure (SRI)} to enhance the limited compute and perception capabilities of standalone vehicles. In particular, several efforts~\cite{he2021vi, he2022automatch, shi2022vips, liu2021livemap} have developed SRI-assisted perception solutions for autonomous driving.
To date, the deployment of SRI is still in its infancy stage. Existing commercial products like roadside units (RSUs)~\cite{datang_gohigh, commsignia_2021}  and experimental prototypes ~\cite{tsukada2020networked, krammer2019providentia, hinz2017designing} are standalone roadside systems that have limited compute and networking capabilities. Moreover, they still have a very low penetration rate and have not been fully validated in real-world deployments. 

To realize the vision of infrastructure-assisted autonomous driving, SRI must address a multitude set of key requirements. First, as a shared infrastructure, SRI should serve a large number of vehicles and diverse autonomous driving applications. For example, it should provide passing vehicles on-road object detection results for informed driving decisions as well as point cloud segmentation results for the purpose of mapping and navigation~\cite{luo2019localization}. Therefore, SRI needs to execute the inferences of {\em multiple concurrent} deep learning (DL) models while meeting stringent real-time requirements. Second, SRI should be capable of both high-bandwidth infrastructure-to-infrastructure (I2I) and infrastructure-to-vehicle (I2V) communication. Data-intensive sensors like LiDAR and cameras have become the {\em de facto} configuration of commercial autonomous driving platforms \cite{Apollo, tusimple_2021,autoware}. 
\blue{These sensors produce high-precision data at a rate of tens of Mbps, which often needs to be shared between SRI nodes. However, to support such a high-bandwidth I2I communication at a large spatial scale, existing technologies like fiber optical Ethernet and 5G cellular networks would incur prohibitively high deployment and operational costs.}
Moreover, although several vehicle-to-everything (V2X) technologies are emerging, they are designed to achieve relatively low data rates \cite{anwar2019physical}. Lastly, to lower the barrier of adoption, the existing traffic infrastructure such as lampposts should be reused or retrofitted as much as possible, which brings various physical and system challenges. For instance, the traffic lampposts typically have a limited power supply (i.e., $\sim 200$\,W in our city), which is challenging to power a fully functional SRI with advanced compute/communication capabilities. 

\blue{A key contribution of this work is to explore a large design space of smart roadside infrastructure and identify a multitude set of key challenges that have not been addressed collectively by existing communication and computing technologies. We present \sysname, the first end-to-end SRI system that is specifically designed for scenarios where the SRI can be deployed at a large spatial scale and operated inexpensively for a long period of time to support autonomous driving.}
Specifically, each \sysname node comprises a low-power single-board edge computer, 802.11ac interfaces, and sensors chosen from three modalities (mmWave radar, LiDAR, and thermal camera), which carefully balances data quality, privacy protection, and power consumption. The communication architecture of \sysname comprises a bi-directional multi-hop I2I network and a downlink I2V broadcast service. 
We exploit the naturally linear topology of the roadside lamppost network and adopt advanced network coding to achieve high-bandwidth and reliable multi-hop I2I communication. A novel I2V broadcast service is designed based on the injector-sniffer mode of 802.11ac, which incorporates lightweight channel switching and empirical measurement-based rate selection for realizing high-bandwidth broadcast to passing vehicles at high speed. 
To tackle the distinct computing challenges encountered in the development of SRI systems, \sysname incorporates a hierarchical task management framework that facilitates desirable load balancing among nodes and enables efficient collaboration. Additionally, we have devised an opportunistic DL task scheduling mechanism to mitigate the impact of resource contention and dynamic system delays arising from multiple concurrent tasks on a single SRI node.

We deployed 18 \sysname nodes on existing outdoor lampposts on campus, operational for over 2 years.
We report experience in the design and deployment of \sysname, as well as extensive experimental results. Our real-world testbed evaluation shows that \sysname can support a diverse set of autonomous driving applications and achieve a $96.1\%$ success rate of application-level data delivery, a $2\times$ improvement over the state-of-the-art baselines. 
Compared to traditional 802.11ac-based approaches, the I2I and I2V communication system of \sysname realizes high-bandwidth and reliable data transmission with $5\times$ improvement in throughput over up to 9 hops and $3\times$ improvement in broadcast bandwidth with multiple vehicles, respectively. 
Our findings and experiences in this work offer key insights into the development and deployment of next-generation smart roadside infrastructure and autonomous driving systems.

%% file: 02-Related.tex
\vspace{-1em}
\section{Related Work}
\label{sec:related}

\myparagraph{Smart roadside infrastructure.} 
Several studies are focused on building real-world SRI with sensors and intelligent devices~\cite{tsukada2020networked, krammer2019providentia, hinz2017designing}. There are also several commercial projects ~\cite{c-its, astri, Friedrichshafen, castermans} on SRI development. However, these efforts are based on small-scale experimental deployments and do not address the comprehensive set of system challenges for autonomous driving. Other works~\cite{Cress2021IntelligentTS, c-its, silva2015deployment, guerna2022roadside, degrande2021c, gopalswamy2018infrastructure, fleck2018towards} focus on roadside unit's software systems and vehicular networks but lack of real-world evaluation and deployment.

\myparagraph{Infrastructure-assisted autonomous driving.} 
Recently, leveraging sensors and compute units installed on roadside infrastructure to assist AVs has received significant attention. Several systems propose to process the sensor data on the infrastructure and provide application-level results to AVs such as object identification~\cite{zhao2019detection,ojala2019novel,zou2022real}, landmark report~\cite{wu2017automatic}, and parking assistance~\cite{cicek2021fully}. A number of studies~\cite{zhao2017cooperative, he2021vi, he2022automatch, shi2022vips, bai2022cyber, arnold2020cooperative, zhang2021emp, liu2021livemap, seebacher2019infrastructure} propose new methods of fusing perceptive information between vehicles or between the infrastructure and the vehicle. These studies focus on specific infrastructure-assisted technologies that shed light on the design objective of \sysname in this work.

\myparagraph{Concurrent DL task execution on cooperative edges.}
Several solutions have been proposed to optimize the performance of concurrent tasks on a single edge node, including on-device DL task scheduling and model compression~\cite{fang2018nestdnn,ling2021rt,kang2021lalarand, yi2020heimdall,ling2022blastnet}. However, they do not address the cooperation among edge nodes, which is essential for fully utilizing the limited resources of the entire edge system. Some studies focus on allocating DL tasks to the edge nodes in a workload-balancing manner~\cite{zeng2020distream,hung2018videoedge}. 
However, these approaches are designed for specific applications and are not applicable to AVs that require high real-time responsiveness.

%% file: 03-Overview.tex
% \vspace{-1em}
\begin{table*}[t]
\centering
\caption{\blue{Cost and performance (Perf.) comparison between \sysname and two alternative paradigms.}}
\vspace{-1em}
\label{tab:cost}
\resizebox{\textwidth}{!}{%
\begin{tabular}{c|cc|cc|cc||cc}
\hline \hline
\multirow{2}{*}{\diagbox{Paradigm}{Module}} & \multicolumn{2}{c|}{I2I} & \multicolumn{2}{c|}{Computing} & \multicolumn{2}{c||}{I2V} & \multicolumn{2}{c}{Cost} \\ \cline{2-9} 
& \multicolumn{1}{c|}{Method}  & Perf.  & \multicolumn{1}{c|}{Method}  & Perf.   & \multicolumn{1}{c|}{Method}  & Perf. & \multicolumn{1}{c|}{Installation} & Operation  \\ \hline
Ethernet+Cloud   & \multicolumn{1}{c|}{Fiber Ethernet}   & \textbf{High} & \multicolumn{1}{c|}{Cloud computing} & \textbf{High} & \multicolumn{1}{c|}{5G V2X}   & \textbf{High} & \multicolumn{1}{c|}{High} & Medium  \\ \hline
5G+Cloud/Edge    & \multicolumn{1}{c|}{5G cellular} & Low/Medium   & \multicolumn{1}{c|}{Cloud computing}  & \textbf{High} & \multicolumn{1}{c|}{5G V2X}  & \textbf{High}  & \multicolumn{1}{c|}{Medium/High} & High \\ \hline
\new{Hybrid 5G/Ethernet+Cloud}    & \multicolumn{1}{c|}{\new{Hybrid 5G/Ethernet}} & \new{Medium}   & \multicolumn{1}{c|}{\new{Cloud computing}}  & \textbf{\new{High}} & \multicolumn{1}{c|}{\new{5G V2X}}  & \textbf{\new{High}}  & \multicolumn{1}{c|}{\new{High}} & \new{Medium/High} \\ \hline
\sysname  & \multicolumn{1}{c|}{802.11ac multi-hop} & Medium   & \multicolumn{1}{c|}{Collaborative edge} & Medium   & \multicolumn{1}{|c|}{802.11ac broadcast} & Medium & \multicolumn{1}{c|}{\textbf{Medium}} & \textbf{Low} \\ \hline \hline
\end{tabular}%
}
% \vspace{-1em}
\end{table*}

\section{Overview}
\label{sec:overview}

\subsection{\blue{Applications}}
\label{subsec:requirements}

\blue{\sysname aims to enable a myriad of infrastructure-assisted autonomous driving applications that provide sensing information to vehicles in real time. Here we highlight three typical applications, each with representative characteristics that pose design requirements for \sysname.}

\myparagraph{\blue{Percpetion sharing.}}
\blue{A crucial application of SRI is to leverage sensors (e.g., LiDARs, etc.) to enhance the perception of autonomous vehicles (AVs). The information shared from the infrastructure to the vehicle can be raw sensor data~\cite{he2021vi, he2022automatch} or computed perception results~\cite{shi2022vips}. AVs can integrate such information from the infrastructure with their own to construct comprehensive scenes and enhance their ability to understand and respond to complex driving environments. As a representative \textit{data-intensive application}, perception sharing imposes the requirement for SRI to process and transmit a substantial volume of data to AVs in real time. For example, sharing LiDAR point clouds with AVs~\cite{he2021vi} requires a data transmission bandwidth over $30\,$Mbps (for a 32-line LiDAR~\cite{velodyne_lidar_2023}).}

\myparagraph{\blue{Traffic monitoring.}}
\blue{Traffic flow monitoring~\cite{zou2022real} offers insights into traffic congestion and potential alternative routes that extend beyond the vehicle's visual range. As an essential requirement, it requires real-time data transmission between infrastructures when vehicles/pedestrians need to be tracked continuously over a period of time. Consequently, the infrastructure system must exhibit satisfying \textit{scalability} and \textit{cost-effectiveness} to support large-scale deployment and coordination. In addition, it underscores the need for \textit{robust communication} between infrastructure nodes.}

\myparagraph{\blue{Accident warning.}}
\blue{Accident warning~\cite{ojala2019novel} can aid AVs in detecting pedestrians and cyclists, especially under conditions of poor visibility or in areas with blind spots. As a life-critical application, it necessitates the infrastructure to deliver warning information to the vehicle timely. Considering that SRI may have to support multiple applications concurrently and mainstream perception-related applications are typically based on compute-intensive deep learning (DL) algorithms, it is essential for the SRI to handle concurrent DL tasks and meet their real-time requirements.}

\subsection{\blue{Design Objectives and Choices}
\label{subsec:design-choice}}

\blue{Drawing upon the requirements derived from applications, we distill several design objectives that are essential in shaping the development and implementation of \sysname, including 1) \textit{high-bandwidth data communication} between infrastructure nodes as well as from infrastructure to vehicles; 2) \textit{efficient inference of multiple concurrent DL models} with stringent realtime requirements; and 3) \textit{low installation/operation costs}, e.g., by reusing/retrofitting the existing traffic infrastructure, for large-scale deployment.}

\blue{We now discuss several design choices based on existing technologies and highlight the key challenges. Table~\ref{tab:cost} shows a comparative analysis of Soar's design choices against two other SRI paradigms.}

\myparagraph{\blue{Ethernet+Cloud.}} 
\blue{A straightforward paradigm of SRI is to connect all infrastructure nodes with specially laid fiber optical Ethernet and offload all computing tasks to the cloud. 
SRI can communicate with vehicles via 5G V2X, the most advanced V2X technology currently available~\cite{naik2019ieee}. This paradigm can deliver robust task execution performance facilitated by high-speed and reliable Ethernet connections as well as the cloud infrastructure. However, it necessitates substantial investment on fiber optical cable installation and cloud servers, particularly in rural regions that lack such network and computing infrastructures. 
In addition, existing V2X technologies (e.g., LTE-V2X) focus on low data rates (up to 25\,Mbps~\cite{anwar2019physical}) and hence cannot handle large data volumes like LiDAR point clouds in real time. 
Moreover, V2X devices on the market remain expensive. For instance, an off-the-shelf Gohigh 5G-V2X transmitter-receiver pair costs over $15k\,$USD~\cite{datang_gohigh}.}

\myparagraph{\blue{5G+Cloud/Edge.}}
\blue{Another alternative paradigm is to offload computation to the cloud while the nodes communicate with the cloud via a 5G cellular network. Compared with the \textit{Ethernet+Cloud} paradigm, this 5G cellular approach demands merely a low-cost 5G interface for each infrastructure node. 
However, the operational expenses of such a system are notably higher due to the service charges of 5G data plans. For example, a LiDAR-equipped SRI node may need to transmit more than a TB of data via 5G during a single day. Furthermore, in areas such as cities with densely deployed roadside SRI nodes, transmitting large volumes of data from SRI to the cloud through a 5G cellular network can incur significant network overhead, thereby adversely impacting application performance. Another approach within this paradigm is to leverage the edge computing capability of 5G cellular networks~\cite{tran2017collaborative}. It typically involves the deployment of dedicated 5G base stations and edge servers specifically designed to execute computing tasks from SRI. Nevertheless, this approach necessitates the installation of numerous additional base stations as well as high-speed network connectivity between them to handle the large amount of data from SRI nodes, which incurs considerably high deployment costs.} 

\myparagraph{\new{Hybrid 5G/Ethernet+Cloud.}}
\new{A hybrid paradigm of SRI leverages Ethernet for high-speed connections between infrastructure nodes. With computing tasks still offloaded to the cloud, it integrates 5G for the communication between infrastructure nodes and the cloud. This paradigm features reliable data transfer and 5G's expansive reach. However, it still incurs significant initial costs for fiber optical Ethernet setup and recurring expenses for 5G data plans.}

\vspace{-1em}
\subsection{System Architecture}
\label{subsec:architecture}

\begin{figure}[t]
	\begin{minipage}[b][][b]{\columnwidth}
		\centering
		\includegraphics[width=0.95\columnwidth]{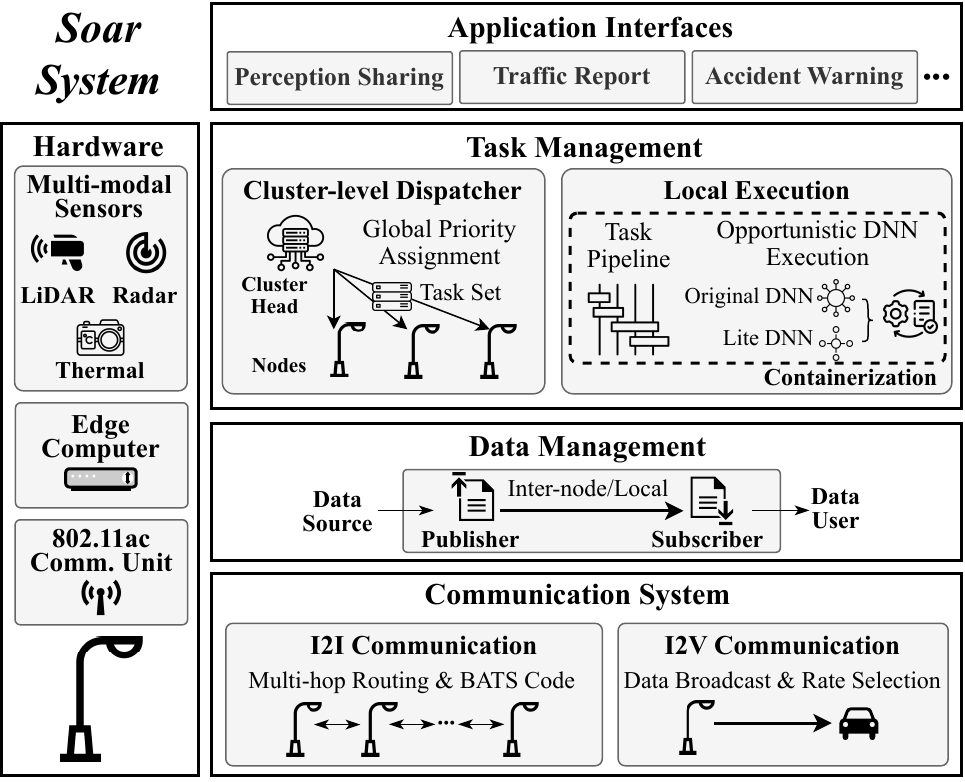}
	\end{minipage}
    \vspace{-2em}
	\caption{The system architecture of \sysname.}
    \vspace{-1em}
	\label{fig:overview}
\end{figure}

\blue{\sysname adopts a systematic design methodology that seamlessly integrates I2I and I2V wireless communication, efficient data/compute task management, and careful power-efficient hardware design. \sysname achieves a remarkably low total power consumption of approximately 70W, making it easily deployable on existing traffic lampposts without requiring extensive upgrades or modifications\footnote {Most operational traffic lampposts have a power supply budget of about $100\sim200\,$W, most of which is used for lighting~\cite{lamppostlight}.}. 
We note that the design choices made by \sysname target specifically the scenarios where SRI needs to be deployed at scale and operated with minimal expense. However, given the highly diverse real-world settings, different technologies would likely coexist and complement each other during the adoption and deployment of SRI.}

Fig.~\ref{fig:overview} shows the system architecture of \sysname. It comprises a series of roadside infrastructure nodes, each equipped with various sensors, an edge computer, and communication interfaces. The \sysname nodes can communicate with each other via a multi-hop wireless network and broadcast to passing vehicles. \sysname adopts a hierarchical task management framework to first allocate DL tasks within a cluster of nodes and further optimizes their concurrent execution on each node at runtime. Moreover, we design a set of application interfaces to support diverse autonomous driving applications.

\myparagraph{Hardware system.} We choose three sensor modalities for \sysname: millimeter-wave (mmWave) radars, LiDARs, and thermal cameras. MmWave radars are used for background tasks such as speed measurement and vehicle counting. LiDARs capture high-resolution yet privacy-preserving 3D information about scenes and objects on the road. Motivated by the fact that \sysname only requires a view of the road, we employ two low-cost solid-state LiDARs with an FOV of around $90^\circ$ facing opposite directions of the road. Compared to traditional omnidirectional LiDARs~\cite{velodyne_lidar_2023}, our design leads to a $10\times$ cost reduction. Thermal cameras are chosen for their night vision capability and the preservation of user privacy, although RGB cameras can also be easily integrated. Each \sysname node equips a single-board edge computer as the main control and compute unit. The communication subsystem of \sysname requires fairly high processing capability to implement reliable high-bandwidth I2V and I2I communication (see \S~\ref{sec:communication}). Therefore, it is implemented using a standalone embedded computer board. Both the sensors and communication unit are connected to the edge computer via a PoE switch. Such a modular design improves both the system robustness and the cost-effectiveness of part replacement. Moreover, each \sysname node achieves a total power consumption of $\sim$70\,W.

 \myparagraph{Communication system.} The communication architecture of \sysname comprises a bi-directional multi-hop I2I network and a downlink I2V broadcast service. 
 \blue{Specifically, we choose to implement I2I communication using a wireless 802.11ac network in a multi-hop manner and adopt advanced network coding to achieve high bandwidth and reliability. 802.11ac can leverage existing roadside infrastructure and consumer-level equipment, lowering upfront and operational costs. The multi-hop approach allows to extend the network's coverage without needing extra infrastructure. We then adopt an I2V communication framework based on the injector-sniffer mode of 802.11ac that focuses on unidirectional downlink broadcasting from \sysname to vehicles. \sysname focuses on passive broadcasting to significantly ease the integration of SRI with existing autonomous driving systems. Applications such as perception sharing and accident warning only need vehicles to receive and process sensor information from the infrastructure, \new{making the passive broadcast model an ideal choice as vehicles in the vicinity of each other usually require identical information about the surroundings.}}

 \myparagraph{Data and task management.} 
For data management, we adopt a publisher-subscriber scheme where \sysname creates a data publisher for each sensor and launches subscribers to fetch data. 
This design decouples the data consumption from sensors, thus achieving the plug-and-play of multi-modal sensors at the software level. Moreover, it prevents sensors from direct access by downstream tasks, which enhances the security of sensor data. 
 
\blue{
Soar adopts a collaborative edge approach \cite{jiang2023coedge} and proposes a novel hierarchical task management architecture that distributes data processing and tasks across local Soar nodes. Thus, Soar can leverage the processing capabilities of local infrastructure nodes so that tasks can be processed closer to the data source to reduce data transmission and latency.}
\sysname nodes are first clustered based on their geographical distribution (e.g., the nodes on the same road section are grouped into the same cluster). The cluster head dispatches tasks within the cluster by jointly considering task urgency and resource availability of nodes. 
Such a cluster-level task allocation approach achieves desirable load balancing among nodes and enables them to collaborate efficiently to run multiple data-intensive applications. 
Moreover, it naturally supports location-awareness of data and compute tasks, such as processing sensor data from nodes around busy intersections and sharing results with vehicles ahead of time. 
In addition to the cluster-level task dispatching, we design an opportunistic DNN execution mechanism for local task execution on each \sysname node. Our key idea is to generate a lite DNN model for each DL task, and then choose one of the versions to execute at runtime, which can accommodate highly dynamic communication bandwidth and computing resources.

\myparagraph{Application interfaces.}
We design a set of application interfaces to efficiently share the sensor data or compute results with vehicles. First, \sysname supports the provision of raw or processed sensor data based on the requirements of different downstream autonomous driving applications. Using LiDAR data as an example, \sysname broadcasts three levels of LiDAR results: the raw point clouds, semantic segmentation results, and object detection results. They can be utilized by mapping~\cite{liu2021livemap}, navigation~\cite{luo2019localization}, and decision-making~\cite{qiu2018avr} tasks on the vehicle, respectively. Moreover, \sysname can also broadcast high-level traffic information such as current traffic congestion level or accident warnings to surrounding vehicles.

%% file: 04-Communication.tex
\section{Communication System}
\label{sec:communication}

\subsection{Multi-hop I2I Network}\label{subsec:I2I}

\begin{figure*}[t]
	\begin{minipage}[b][][b]{1.25\columnwidth}
		\centering
		\includegraphics[width=\columnwidth]{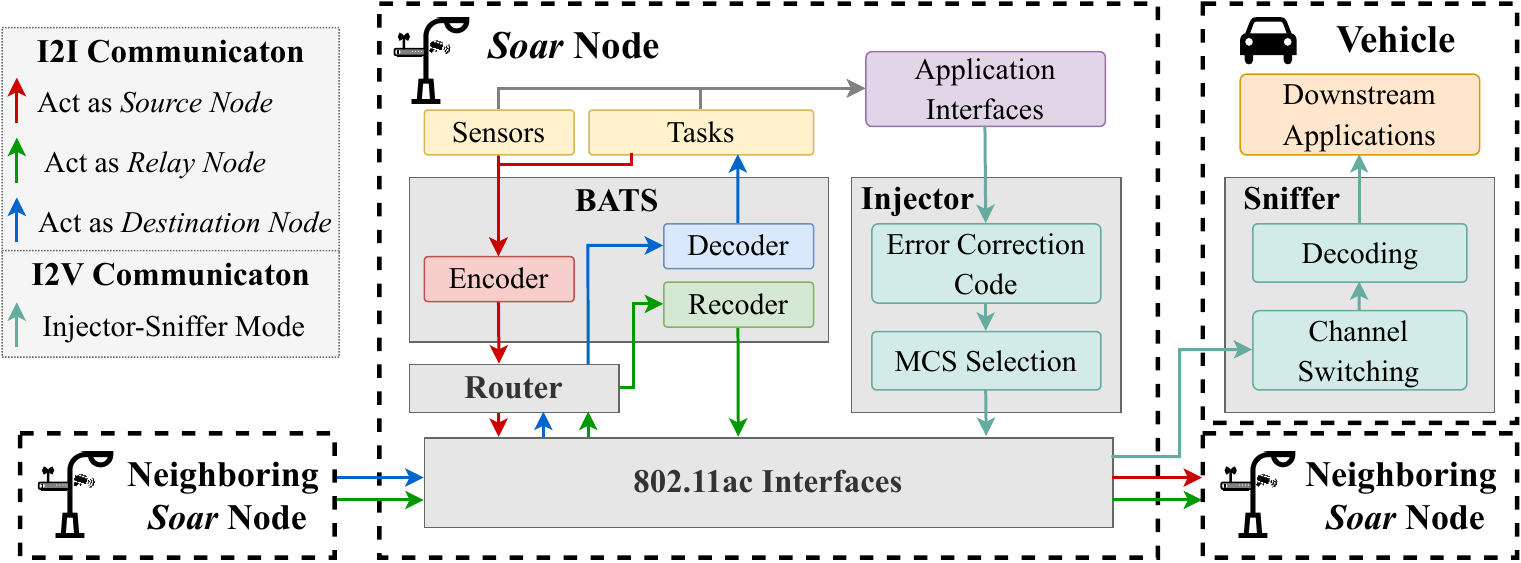}
    \vspace{-1.5em}
	\caption{Communication system architecture of \sysname.}
    \vspace{-1.5em}
	\label{fig:communication_system}
    \end{minipage}
    \hspace{1em}
    \begin{minipage}[b][][b]{.75\columnwidth}
		\centering
		\includegraphics[width=0.98\columnwidth]{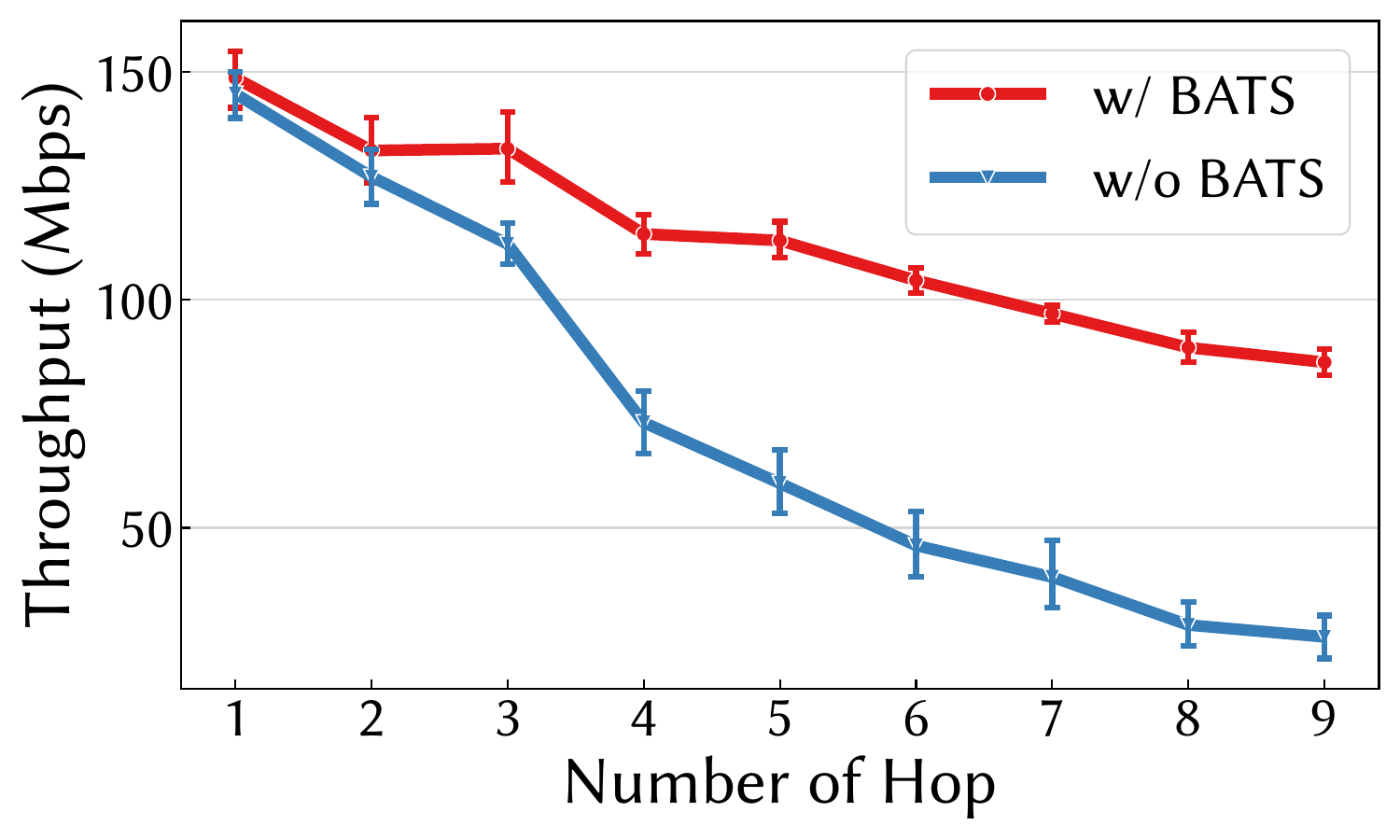}
    \vspace{-1.5em}
	\caption{Multi-hop TCP performance w/ and w/o BATS code.}
    \vspace{-1.5em}
	\label{fig:bats_result}
    \end{minipage}
\end{figure*}

\begin{figure*}[!htb]
    \centering
    \begin{minipage}[b][][b]{.50\columnwidth}
		\centering
		\includegraphics[width=\columnwidth]{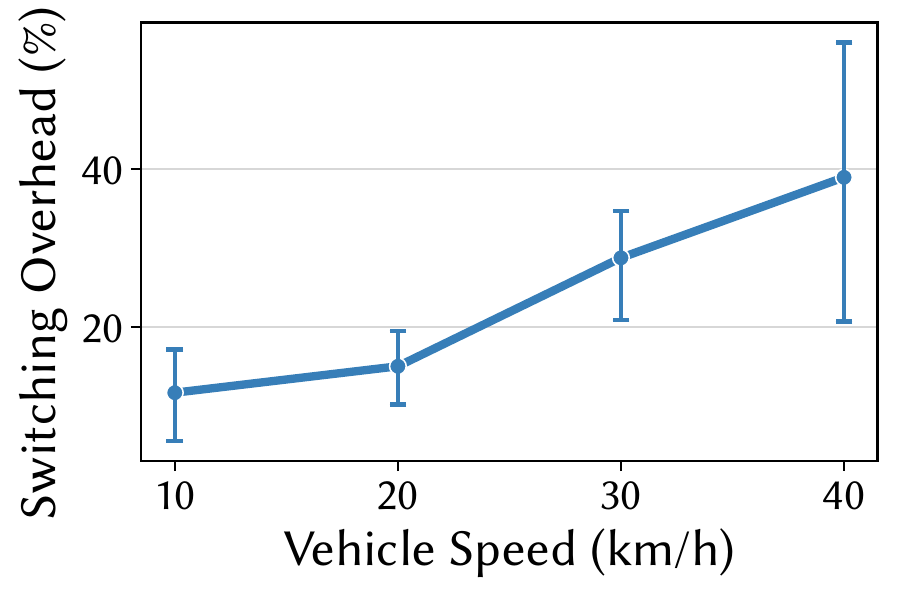}
        \vspace{-2em}
	    \captionof{figure}{802.11ac switching overhead.}
        \vspace{-1em}
	\label{fig:switch_stats}
    \end{minipage}
    \hspace{.5em}
    \begin{minipage}[b][][b]{1.50\columnwidth}
        \centering
        \begin{subfigure}[b]{0.32\linewidth}
            \includegraphics[width=\textwidth]{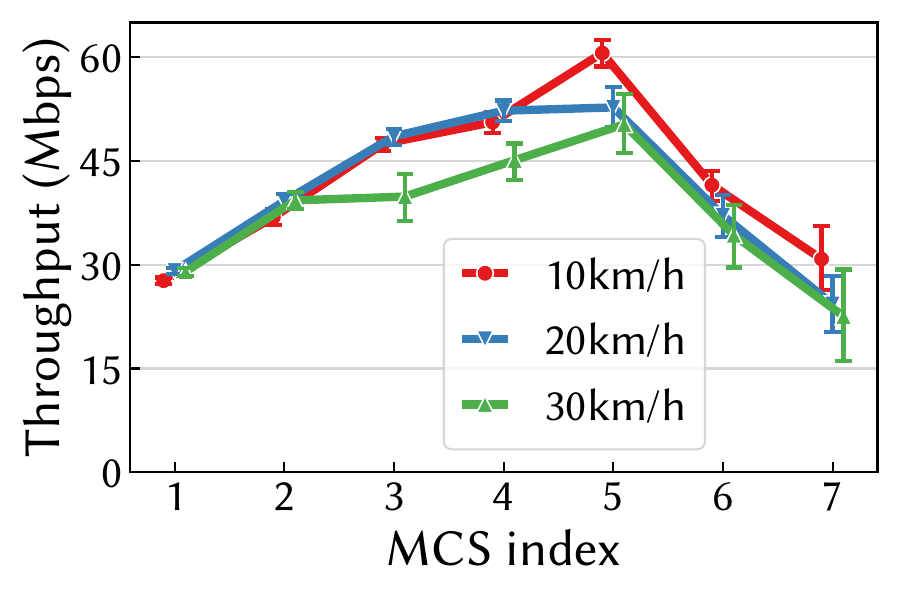}
            \vspace{-2em}
            \caption{Campus road.}
            \label{subfig:mcs_campus}
        \end{subfigure}
        \hfill
        \begin{subfigure}[b]{0.32\linewidth}
            \includegraphics[width=\textwidth]{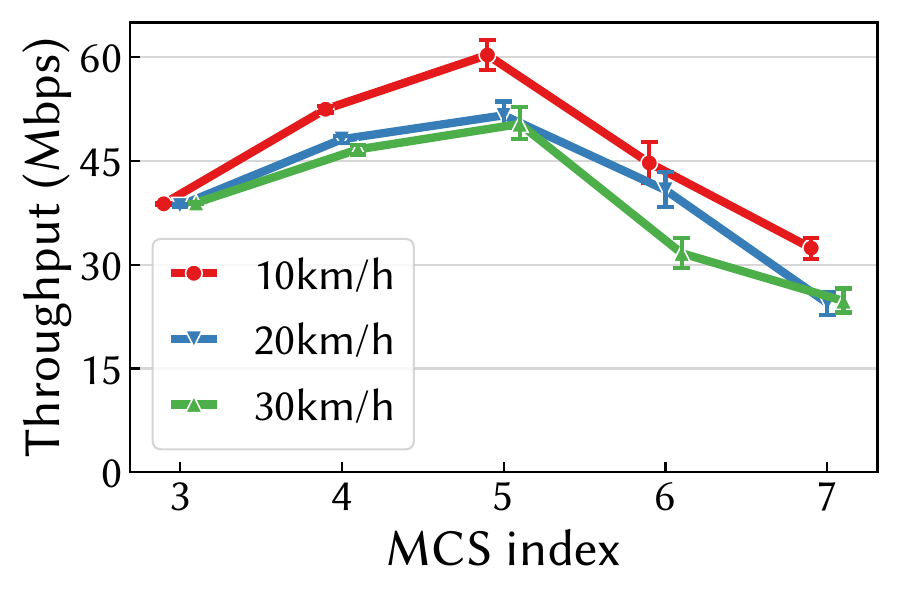}
            \vspace{-2em}
            \caption{Community road.}
            \label{subfig:mcs_community}
        \end{subfigure}
        \hfill
        \begin{subfigure}[b]{0.32\linewidth}
            \includegraphics[width=\textwidth]{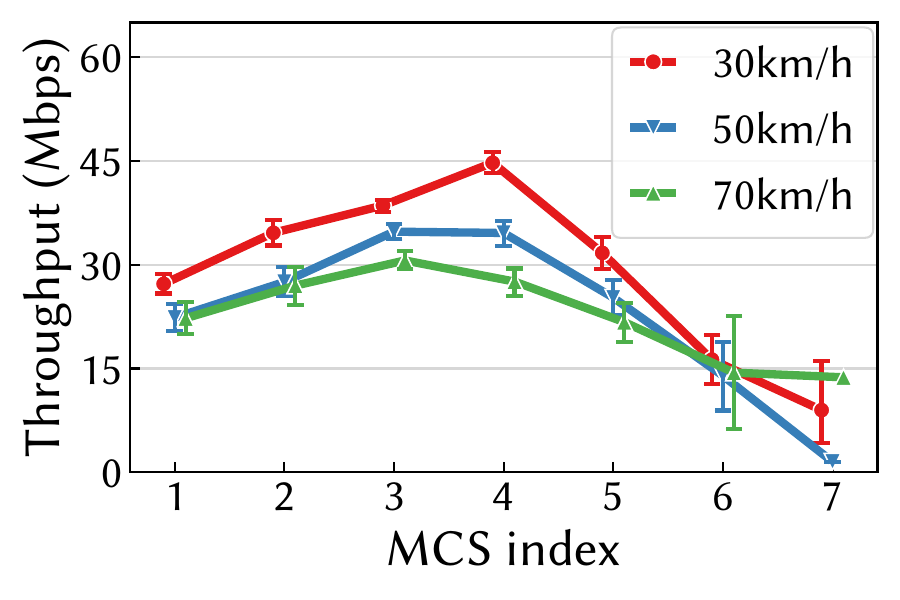}
            \vspace{-2em}
            \caption{Public road.}
            \label{subfig:mcs_public}
        \end{subfigure}
        \vspace{-1em}
        \captionof{figure}{Measurements of MCS performance in different road environments where 802.11ac with 2 spatial-stream is applied.}
        \vspace{-1.5em}
        \label{fig:mcs_measure}
    \end{minipage}
\end{figure*}

A key design objective of \sysname is to achieve high-bandwidth multi-hop I2I communication, i.e., about $100$\,Mbps over up to $10$ hops within $500$\,meters. This allows the \sysname nodes located on the same road section to achieve efficient coordination and load balancing. As depicted in \S~\ref{sec:overview}, \sysname leverages a cluster-level task allocation approach to overcome the limited compute resources on each node. Such task offloading requires high-bandwidth and reliable data transmission among \sysname nodes. For instance, a typical LiDAR produces point clouds at a rate of more than $30$\,Mbps~\cite{velodyne_lidar_2023}. 
\blue{As discussed in \S~\ref{subsec:design-choice}, we opt to employ off-the-shelf 802.11ac to implement I2I communication of \sysname for its wide availability and cost-effectiveness.}
The left part of Fig.~\ref{fig:communication_system} illustrates our I2I communication architecture. 

\myparagraph{Linear network topology and multi-hop routing.} 
\blue{The mesh topology is widely adopted in multi-hop wireless ad hoc networks~\cite{dsr,aodv}. However, it is well known that the throughput of wireless mesh networks is low in real-word settings due to severe interference and channel contention. The design of \sysname addresses this challenge by exploiting the naturally linear topologies of roadside nodes.}
Specially, we implement a multi-hop routing strategy where each node has two bridged 802.11ac interfaces operating in AP and STA modes, connecting to previous and next nodes. 
\new{This design offers greater scalability compared to Ethernet or 5G cellular networks in terms of cost-effectiveness and deployment flexibility.}
In case of link failures, nodes automatically attempt to establish a connection with a further node.
We adopt such a semi-fixed routing strategy instead of fully dynamic multi-hop routing protocols ~\cite{bicket2005architecture,perkins1994highly,de2003high} 
because the link between adjacent \textit{Soar} nodes typically have a line of sight and short distance (e.g., $30\sim50$\,m~\cite{berriman_2019, pusat2002}), resulting in good link quality. 

\myparagraph{Network coding.} 
Multi-hop wireless networks often suffer from significant packet loss~\cite{al2011tcp}. While various link layer techniques, such as network coding~\cite{network_coding_1,network_coding_2} and fountain codes~\cite{fountain_1,fountain_2}, have been proposed to enhance multi-hop reliability, the high compute and storage costs limit their applicability to resource-constrained edge platforms~\cite{yang2014bats,zhang2016fun}. 
\new{To tackle this challenge, we employ the Batched Sparse (BATS) code~\cite{bats_code}, an advanced network coding approach that addresses the packet loss in linear multi-hop networks with extremely low power and storage consumption. We exploit it to optimize end-to-end communication throughput and relieve heavy retransmission overhead.}
Specifically, BATS code features an outer code and an inner code, with the outer code being a matrix generalized fountain code generating numerous batches for high capacity. 
During the multi-hop transmission, an inner code derived from random linear network coding~\cite{RLNC} is applied to the batched packets at each relay hop, realizing persistent reliability with low overhead.
Though several studies~\cite{yang2014bats,bats_impl_1} have demonstrated the performance of BATS code in simulations or lab settings, \sysname is the first system to apply BATS codes in real-world I2I communication.
\blue{We implement BATS codes at the network layer to be compatible with the existing network stack, which is transparent to traditional socket and transport layer protocols. We utilize \textit{Netfilter}~\cite{netfilter} to capture IP packets and apply different coding depending on their destinations. Fig.~\ref{fig:bats_result} shows \textit{iperf3}~\cite{iperf} results of a 10-node wireless linear network. With BATS codes, TCP throughput can achieve above $100$\,Mbps over 6 hops and $90$\,Mbps over 9 hops, an up to $5\times$ improvement compared to the baseline without BATS codes.}

% \vspace{-1em}
\subsection{I2V Broadcast}\label{subsec:I2V}

\sysname needs to transmit large volumes of data to passing vehicles for downstream autonomous driving tasks, such as providing LiDAR point clouds for 3D perception, which demands $\sim30$\,Mbps bandwidth~\cite{he2021vi}. 
\sysname aims to achieve a communication bandwidth of $\sim50$\,Mbps between SRI and multiple vehicles. 

\myparagraph{Injector-sniffer-based high-bandwidth broadcast.} 
\blue{As discussed in \S~\ref{subsec:design-choice},} we propose an 802.11ac-based I2V communication framework for high-bandwidth, cost-effective data broadcast from each \sysname node to nearby vehicles. 
\new{Our design focuses on the unidirectional downlink broadcasting and utilizes the injector-sniffer mode~\cite{injector,sniffer} to achieve high data rates without link establishment.
\sysname provides either raw sensor data or detection results (e.g., bounding boxes), which can support a range of autonomous driving tasks that rely on SRI data to enhance perception and reliability.}
The \sysname node acts as an injector, transmitting raw wireless packets with headers into the air, while vehicles sniff packets using onboard 802.11ac receivers.
\new{This broadcast design without vehicle feedback makes \sysname highly scalable with respect to the number of vehicles in the network.}
The right part in Fig.~\ref{fig:communication_system} shows the I2V communication pipeline. \sysname node first encodes data with error correction, then modulates it using an empirical rate selection scheme, and finally transmits on a pre-assigned channel. Vehicles decode packets, recover original data for applications, and determine the channel to receive messages based on location or signal strength.

\myparagraph{Error correction code.}
Without vehicle feedback, our passive data broadcast design may experience severe packet loss, hindering goodput for downstream applications. \blue{To tackle this challenge, we apply a systematic random linear code~\cite{gallager1962low} for error correction, which generates multiple check packets on application layer to ensure high data delivery reliability.}

\myparagraph{Measurement-based rate selection.}
After applying error correction, \sysname modulates data for transmission. A challenge in designing the modulation and coding scheme (MCS) is selecting the optimal rate for reliable high-throughput transmission. 
Existing works on the multicast rate selection rely on receiver feedback and thus are incompatible with our passive design. 
\blue{Our extensive real-world experiments show that there exists a trade-off between modulation rate and packet loss, depending on the environmental dynamics. \new{Fig.~\ref{fig:mcs_measure} illustrates that there exists an optimal MCS that demonstrates consistent performance under different vehicle speeds but varies with the road environments. This is because the impact of the environment (i.e., road shape, trees, buildings, etc.) dominates the performance of I2V communication under different MCSs, while the impact of low urban speeds (i.e., $<70$ km/h) is not significant.} Motivated by this observation, \sysname adopts an empirical MCS rate selection scheme where each node is configured with a fixed rate based on installation-phase measurements.}

\myparagraph{Lightweight channel switching.}
Unlike 802.11ac incurring significant reassociation overhead at high vehicle speeds (Fig.~\ref{fig:switch_stats}), our design only requires vehicles to switch sniffing channels to receive data from \sysname continuously. With 24 non-overlapping channels in the 5GHz band, each \sysname node can be assigned a different channel without inducing the interference among the same cluster.
\blue{When a strong interference level (e.g., caused by nearby Wi-Fi, 5G signals, etc.) is detected on the assigned channel, \sysname node will switch to another free channel.}
\blue{Each \sysname node will periodically broadcast the channel assignments for its cluster on a pre-fixed control channel (e.g., channel 36). Passing vehicles first listen to the control channel for channel assignments, and then switch channels during movement based on their own and nearby \sysname nodes' locations}
\footnote{The locations of \sysname nodes are easily accessible by vehicles since they are installed on existing lamppost infrastructure. \sysname nodes may also obtain their own locations through GPS and broadcast to vehicles.}.

\myparagraph{\blue{Security Issues.}}
\blue{Without association and uplink from vehicles to the infrastructure, our passive I2V broadcast design may be vulnerable to security issues like malicious data injection. To mitigate such attacks, we employ private key encryption~\cite{diffie2022new} to create a digital signature within data packets. Certified vehicles can obtain a public key, which is used to verify the signature upon data receipt to ensure security.}

%% file: 05-Computing.tex
\begin{figure}[t]
	\begin{minipage}[b][][b]{\columnwidth}
		\centering
		\includegraphics[width=1\columnwidth]{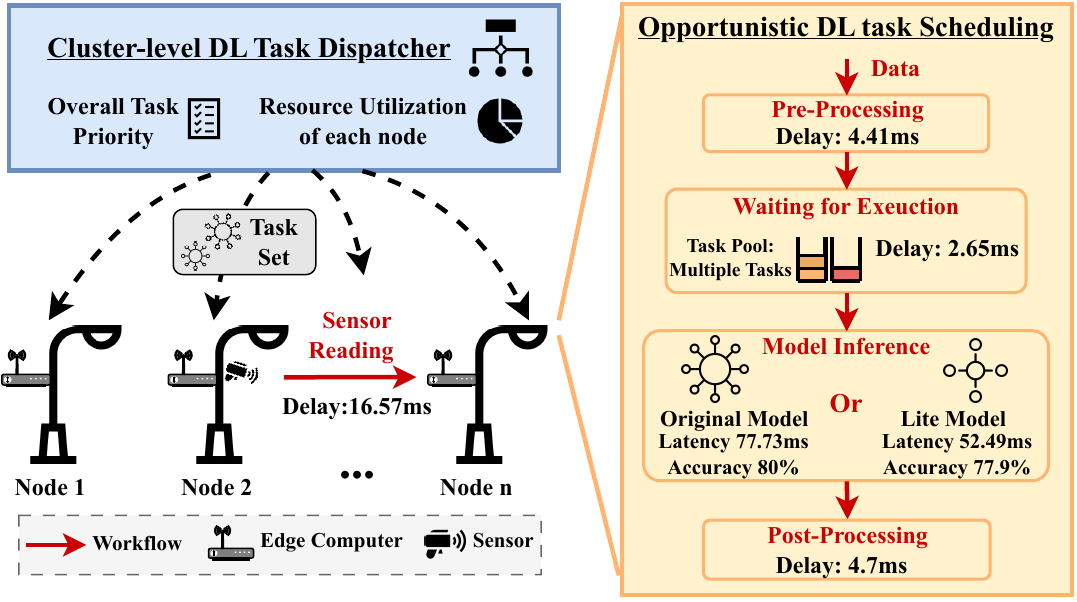}\\[-1ex]
	\end{minipage}\vspace{-1em}
	\caption{The task management framework of \sysname.}
    \vspace{-1.5em}
	\label{fig:computing-global}
\end{figure}

\section{Task Management}
\label{sec:computing}

A key challenge of \sysname's design is supporting executing multiple real-time DL tasks concurrently on resource-constrained edge platforms. We propose a novel task management framework that can efficiently dispatch DL tasks among multiple \sysname nodes in a collaborative manner. Our design is motivated by the following characteristics of roadside infrastructure nodes. First, the DL tasks on different nodes are highly diverse, due to different road sections and heterogeneous hardware/sensor configurations. For instance, the \sysname nodes at the crossroads are typically installed with more types of sensors for complex tasks such as vehicle tracking and pedestrian detection. Second, the communication bandwidth and compute resource available on each \sysname node are highly dynamic. For instance, traffic spikes during rush hour may lead to a fluctuation in communication bandwidth.

\blue{To address these challenges, as shown in Fig.~\ref{fig:computing-global}, our task management framework is based on a two-tier clustering structure, where the \sysname nodes can be naturally clustered based on road sections or geographic locations. First, \sysname employs a cluster-level task dispatcher, which is responsible for balancing task workloads in a cluster by jointly considering task priorities/deadlines, resource availability, and geographical positions of nodes. Second, \sysname includes a local opportunistic task scheduler to control the execution of multiple DL tasks on a single \sysname node, which aims to optimize the real-time performance of concurrent DL tasks under runtime system dynamics.}

\vspace{-1em}
\subsection{Cluster-level DL Task Dispatching}
\label{subsec:global}

We design a cluster-level task dispatcher for SRI based on both overall task priorities and resource utilization of each \sysname node.
Our goal is to maximize the number of total deployed tasks while meeting their real-time requirements, as shown in Eq.~\ref{eq:computing_goal}. $Priority(m)$ denotes the task priority weight of task $\tau_{m}$. $N(\tau_{m},n)$ equals 1 if the task $\tau_{m}$ is deployed on the Node $n$, else it equals $0$. $T^{E2E}(\tau_{m},n)$ denotes the end-to-end delay of the task $\tau_{m}$ executed on Node $n$, which is defined as the total delay between the launch and the completion of a task. \blue{ $T^{E2E}$ also includes the communication time, which is estimated based on the periodically measured bandwidth.} $DDL(\cdot)$ is the expected completion time (i.e., deadline) of each task, which is not less than the trigger period $Period(\tau_{m})$ of the task $\tau_{m}$. 
% \vspace{-1em}
\begin{equation}
\begin{aligned}
\max_{\forall m} \quad & \begin{matrix}\sum_{m=1}^{M}\sum_{n=1}^{N}Priority(m)\times N(\tau_{m},n)\end{matrix}\\
% \textrm{s.t.} \quad &  MR(\tau_{m},n)\leq \sigma_{m}
\textrm{s.t.} \quad &  T^{E2E}(\tau_{m},n)\leq DDL(\tau_{m}), \begin{matrix}\sum_{n=1}^{N}N(\tau_{m},n) \leq 1\end{matrix}
\label{eq:computing_goal}
\end{aligned}
\setlength\belowdisplayskip{2pt}
\end{equation}
% \vspace{-1em}
An exhaustive search of the problem formulated in Eq.~\ref{eq:computing_goal} has a complexity of $O(MN)$. We adopt an efficient heuristic as follows.
First, we sort the task pool from high to low priority according to the ascending order of $1/deadline$.
We also map the task priorities to a sequence of power of 2. After this mapping, the task with the longer relative deadline is prioritized as the lower priority $2^{0}$. In this way, the priority weight $Priority(m)$ of tasks with high priorities can dominate those with low priorities. 
Then, we determine the deployed node for each task from high to low priority weight. 

For each task, we search for its deployment node from its source node to other adjacent nodes based on the physical proximity. If the time constraint in Eq.~\ref{eq:computing_goal} is met, we search for a \sysname node that has the lowest resource utilization to deploy the task $\tau$. The resource utilization of a Node $n$ can be estimated by the sum of the task laxity of all the tasks on this node, where the laxity for task $\tau_{m}$ is defined as $T^{total}(\tau_{m},n)/Period(\tau_{m})$. 
 $T^{total}(\tau_{m},n)$ denotes the total execution time of the task $\tau_{m}$ on Node $n$, which contains sensor reading, task pre/post-processing and model inference.

\vspace{-1em}
\subsection{Opportunistic DL Task Scheduling}
\label{subsec:local}

Due to severe resource constraints and significant dynamics in the real world, it is challenging to support multiple real-time DL tasks concurrently on the edge platform of the \sysname node. As shown in Fig.~\ref{fig:computing-global}, an end-to-end DL task contains sensor reading, pre-/post-processing, blocking, and model inference. Significant dynamics in the wild such as power surges will cause unpredictable delays in sensor reading. Moreover, resource contention between different system processes (e.g., sensor reading, CPU/GPU processing etc.) will cause dynamic blocking delays. To address these challenges, \sysname adopts an opportunistic execution mechanism to mitigate the impact of fluctuating system delays on the inference of multiple concurrent tasks.

Our main idea is to generate a {\em lite model} for each DL task, which can better adapt to unpredictable resource availability at runtime. 
The lite model can be a compressed DNN model or a small model that distillates knowledge from the original model. 
\blue{However, the lite model achieves a desirable low latency at the expense of accuracy. To mitigate the potential accuracy loss, we design an online opportunistic scheduling algorithm, which executes the lite models only when the time constraints cannot be met, thus minimizing the impact on task accuracy.}
By exploiting such model compressibility, \sysname can reduce the resource demand of each inference according to the runtime condition. \sysname preloads both the original and the lite models during initialization to avoid dynamic loading delays. Our design can also work with existing storage optimization methods such as weight sharing~\cite{ling2021rt,fang2018nestdnn} to reduce the storage overhead of lite models.

Given a task set allocated by the cluster head, we split each task into pre-/post-processing and model inference. These task segments are then assigned to different processes and executed in a pipelining manner. \blue{This design enables full utilization of both CPU and GPU resources by executing pre/post-processing for one task alongside the inference of other tasks in parallel.}
At runtime, we first determine which model to be executed according to the urgent level of each model inference. Specifically, \sysname chooses the model inference job with the latest deadline in the queue for execution. 
\sysname then determines the DNN model (i.e., the original or lite model) to be executed according to the remaining time of the current and the next inferences. We estimate the completion time of each model inference through the measurement results from offline profiling. Meanwhile, \sysname checks one more job in the queue. If there is not enough remaining burst time for the next urgent inference, \sysname also uses the lite model for the current inference. \blue{The jobs that miss the deadline are dropped before execution to prevent the delay accumulation.}

%% file: 06-Deployment.tex
\begin{figure*}[t!]
    \begin{minipage}[b][][b]{2\columnwidth}
	\centering
	\includegraphics[width=\columnwidth]{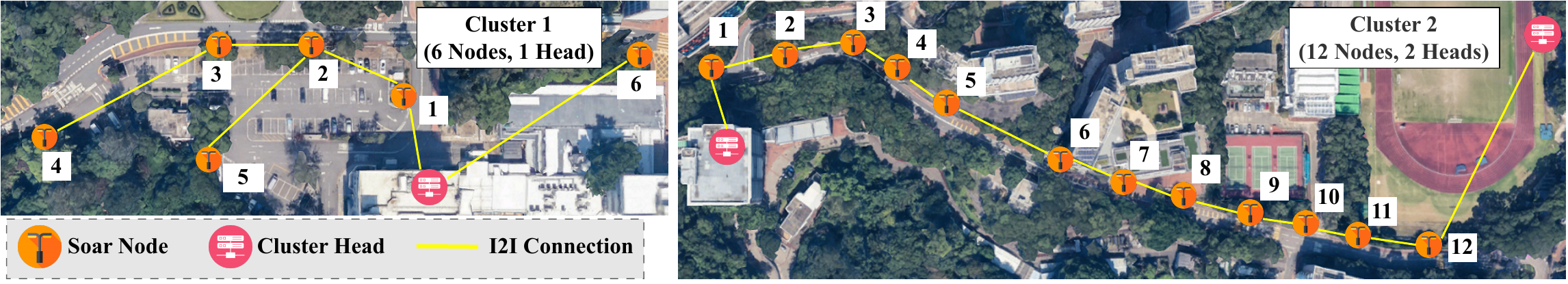}
    \end{minipage}
    \vspace{-1em}
    \caption{Deployment of two \sysname clusters on campus.}
    \vspace{-1.5em}
    \label{fig:deployment_map}
\end{figure*}

\begin{figure}[t!]
    \begin{minipage}[b][][b]{\columnwidth}
        \centering
        \begin{subfigure}[b]{0.49\linewidth}
        \centering
            \includegraphics[width=.9\textwidth]{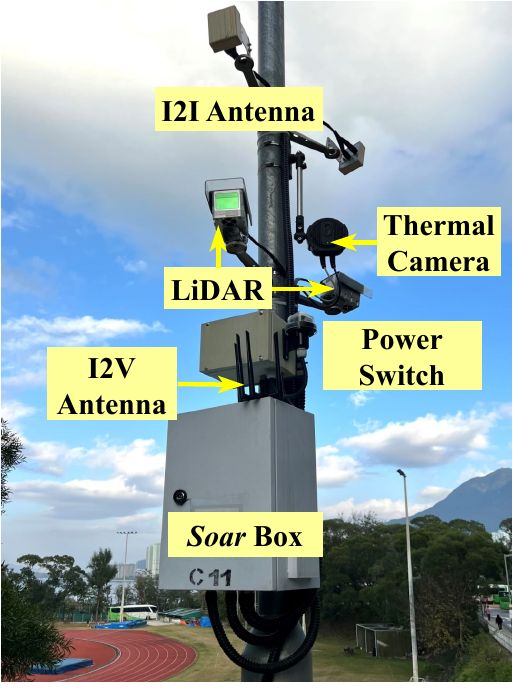}
            \vspace{-.75em}
            \caption{The prototype of \sysname.}
            \label{subfig:prototype}
        \end{subfigure}
        % \hspace{.1em}
        \begin{subfigure}[b]{0.49\linewidth}
         \centering
            \includegraphics[width=.9\textwidth]{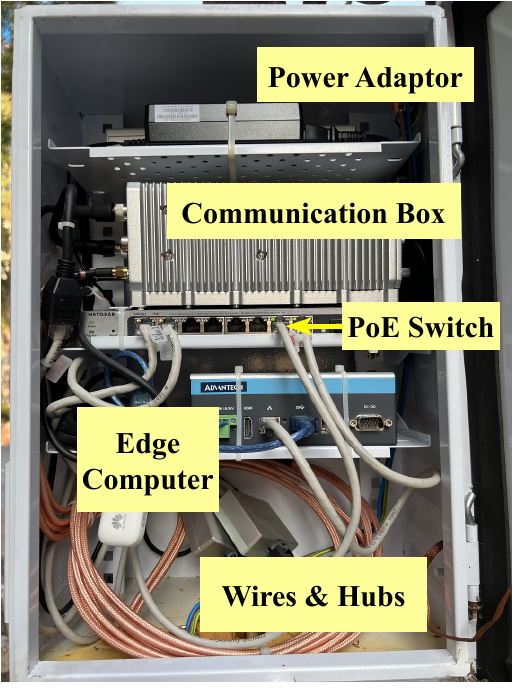}
            \vspace{-.75em}
            \caption{\sysname box layout.}
            \label{subfig:layout}
        \end{subfigure}
    \end{minipage}
    \vspace{-2.5em}
    \caption{Hardware system of the \sysname.}
    \vspace{-2em}
    \label{fig:hardware}
\end{figure}

\section{Implementation and Deployment}
\label{sec:deployment}

\subsection{System Implementation}
\label{subsec:implementation}

We deployed 18 \sysname nodes on existing lampposts in two clusters on our campus, as shown in Fig.~\ref{fig:deployment_map}. The first cluster consists of 6 nodes around a parking lot, with a server inside a building serving as the cluster head. The second cluster has 12 nodes along the main campus road and two cluster heads in buildings. The two clusters cover 0.3 km and 0.5 km of road, respectively. Our campus testbed has received approval from the Institutional Review Board (IRB).

Fig.~\ref{subfig:prototype} shows the prototype of \sysname implementation. Each \sysname node is mounted at the bottom of a lamppost, whose components and layout are shown in Fig.~\ref{subfig:layout}. 
The system uses an NVIDIA Jetson TX2 computing board~\cite{tx2} in a waterproof enclosure, and other devices such as the communication box and sensors are connected to it via a PoE switch.  
\blue{We implement the communication system in a separate industrial single-board computer whose hardware is modified to support multiple 802.11ac network cards. To support I2V broadcast, we modified the RTL8814AU~\cite{8814au} driver to enable customized packet injection.}
Most nodes are equipped with a Teledyne FLIR thermal camera~\cite{TeledyneFLIR} and two Livox Horizon LiDARs~\cite{livox}, and some also have a TI mmWave radar~\cite{iwr6843isk}. Two or three directional antennas are mounted on the top, pointing at adjacent nodes, and a set of I2V antennas is installed below, facing the road. \blue{Each node costs $10.2k$ USD approximately, with the sensors costing around $2.75k$ USD and other components costing $1.95k$ USD.}
We implement the sensor data management module based on ROS2~\cite{ROS2}, which offers various sensor drivers and APIs for realizing the publisher-subscriber mode data management.
The task management module is implemented in KubeEdge~\cite{kubeedge}, 
\blue{which allows the cluster head to efficiently dispatch task codes and their environmental dependencies to each node through the containerization technique.}
\new{To periodically estimate the I2I communication bandwidth between \sysname nodes during cluster-level task dispatching, we use \textit{iperf3}~\cite{iperf} to assess bandwidth on each link every 10 minutes, which incurs negligible overhead.}

% \vspace{-1em}
\subsection{Deployment Experience}
\label{subsec:experience}

\myparagraph{Installation considerations.}
To save lighting power during day time, the lampposts on campus are powered on/off automatically by a solar timer-controlled switch on the circuit bus.
To power \sysname nodes from lampposts continuously, we removed the switch on the bus and modified the power supply on each lamppost into two parts, one nonstop power supply for each \sysname node and a solar timer-controlled one for lighting. 
Moreover, by calculating the load-bearing constraints in extreme weather conditions, we limit the installation height below $5$\,m, the weight below $30$\,kg, and the lantern windage area below $0.12$\,$\text{m}^2$. 
As a result, we installed the antennas and sensors above $3$ meters on the lamppost due to their light weight (i.e., around $5$\,kg), which avoids any occlusion to the view of the road.

\myparagraph{System durability and robustness.}
We also optimized the durability and robustness of \sysname for in-the-wild deployment according to more than two years of operational experience. 
We added a shelf to hold the box of computing devices and store the waterproof cables at the bottom, which can effectively avoid water accumulation over time. 
Moreover, during the long-term operation of \sysname, we observed that the lamppost power supply could occasionally fall short of the operational power requirement of \sysname, causing a system reset. 
This is due to either dynamic grid conditions or the high instantaneous power draw at the startup of hardware devices. 
For example, Jeston AGX Xavier~\cite{xavier} consumes 65\,W of power at startup but only up to 30\,W at runtime. 
This problem can be addressed by adding an uninterruptible power supply, which has been validated on our system.

\myparagraph{Edge software architecture.}
To realize the containerized software deployment, we initially chose Kubernetes (K8S)~\cite{kubernetes} due to its support for container management. 
However, during our deployment, we found various issues, such as the unstable connection between \sysname nodes and the cloud server (i.e., via TCP socket). In this case, the container deployment from the server to the \sysname nodes can be terminated or even forcibly removed by K8S. Therefore, we switched to KubeEdge~\cite{kubeedge} as our development framework, enabling the autonomous edge operation even during disconnection from the cloud. 
Furthermore, the KubeEdge-based implementation occupies only up to 80\,MB of memory, a 60\% reduction compared to the K8S-based implementation.

%% file: 07-1-Exp-Setup.tex
\subsection{Evaluation Methodology}\label{subsec:methodology}

\begin{figure}[t]
	\begin{minipage}[b][][b]{.8\columnwidth}
		\centering
		\includegraphics[width=\columnwidth]{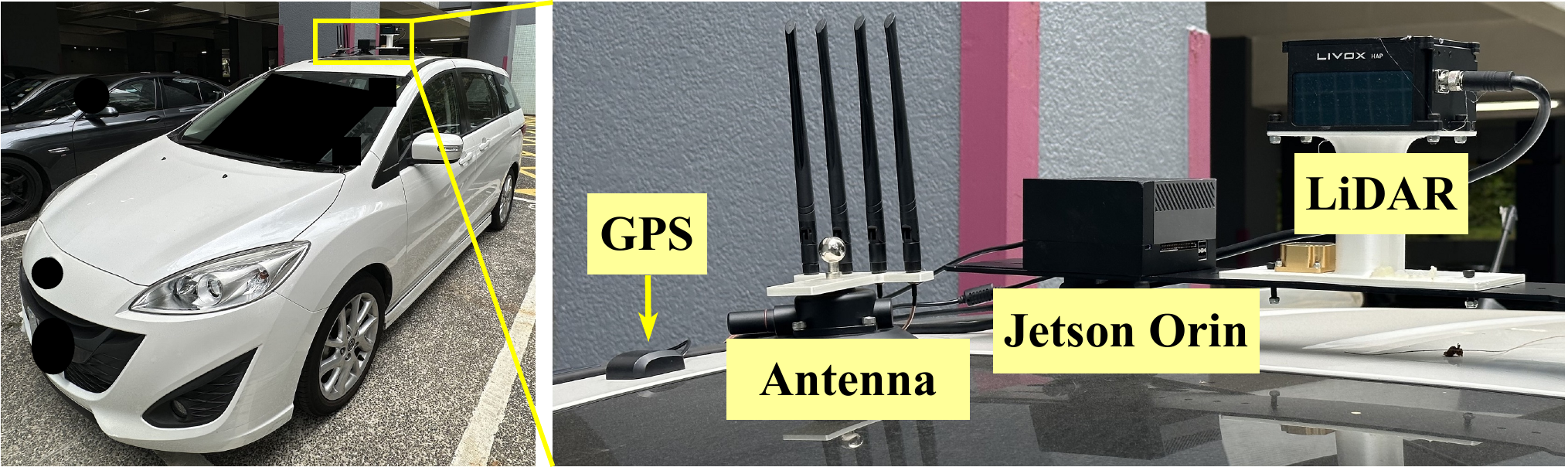}\\[-1ex]
	\end{minipage}
    \vspace{-1em}
	\caption{The test vehicle.}
    \vspace{-2em}
	\label{fig:test_vehicle}
\end{figure}

\myparagraph{Field studies and self-collected dataset.} We conduct extensive experiments based on the two clusters of \sysname nodes on campus (see \S~\ref{sec:deployment}). We drive a test vehicle equipped with a LiDAR, a GPS, a four-antenna array, and a Jetson Orin (see Fig.~\ref{fig:test_vehicle}) to continuously receive data from \sysname nodes. 
To simulate the presence of other vehicles receiving data from \sysname, we randomly placed some fixed 802.11ac devices as pseudo vehicles receiving data from nearby \sysname nodes to simulate different traffic conditions. Moreover, we collect infrastructure- and vehicle-side LiDAR point cloud datasets ($4250$ frames in total) and manually annotate them to train models for evaluating application performance on \sysname. 

\myparagraph{Application implementation.} \blue{We implement three typical autonomous driving applications that require executing DL tasks on \sysname, which are shown in Table~\ref{tab:outdoor_para}. 
These applications include LiDAR-based perception sharing, RGB camera-based traffic monitoring, and thermal-camera-based jaywalk warning. 
Due to the privacy concerns of the campus testbed, we deploy a virtual camera in our data management framework that streams KITTI data \cite{Geiger2012CVPR}. 
For both traffic monitoring and jaywalk warning, we choose YOLOv5s-based and YOLOv5n-based models as the original and lite models.
These models are trained using the KITTI dataset and the Teledyne FLIR ADAS dataset \cite{TeledyneFLIR}, respectively.
In the perception sharing, we compress the original PointPillars \cite{lang2019pointpillars} with width/depth scaling methodology \cite{tan2019efficientnet} to generate a lite model, and train them with our self-collected dataset.
\new{These three applications run concurrently on each node for all the experiments, and the results are transmitted to the vehicle.}}

To further evaluate the benefits of \sysname for AVs, we implement two typical infrastructure-assisted perception fusion applications: point cloud registration and LiDAR perception extension. 
In point cloud registration, the vehicle receives raw LiDAR point clouds and aligns them with its own point clouds. The LiDAR perception fusion is the downstream application of perception sharing, where the vehicle receives the object detection results and fuses them into its view.

\myparagraph{Metrics.}
For evaluation of the communication system of \sysname, we are interested in \textit{throughput}, \textit{switching overhead}, and \textit{packet delivery ratio (PDR)}. The switching overhead is measured by calculating the fraction of time spent on switching the channel during receiving data from a \sysname node. To evaluate the task management performance, we quantify the real-time performance achieved by our framework using the \emph{deadline missing rate}, a widely-used metric for real-time performance evaluation \cite{bateni2018predjoule, ling2021rt}. Specifically, the deadline missing rate contains the job drop and exceed ratios, which indicate the ratios of jobs dropped because of overdue and failing to meet their deadlines, respectively. We also statistic the \textit{average end-to-end (E2E) delay} (defined in \S~\ref{sec:computing}) of each job to quantify the execution efficiency of the DL tasks. Moreover, to focus on the overall performance of \sysname in the applications, we define a \textit{failure rate} to indicate the percentage of cases where the system failed to deliver application data to the vehicle. The failure cases include failing to execute in time and failing to transmit the results to the vehicle. 

\myparagraph{Baselines.} We compare both our I2I and I2V communication with the traditional 802.11ac approaches. Specifically, we evaluate the I2I communication performance w/ and w/o BATS code. Besides, the I2V communication of \sysname is compared with traditional 802.11ac broadcast and unicast modes. For comparison of our task management framework, we implement a task execution mechanism called Local-EDF which executes the DL tasks locally on their source nodes and adopts the Earliest Deadline First (EDF) scheduling policy for concurrent DL task execution. For the evaluation of the overall performance of \sysname, we combine the 802.11ac unicast approach and the Local-EDF task management scheme as the baseline. \blue{We also compare \sysname with three cloud-based baselines. In addition to the settings outlined in Table~\ref{tab:cost} (referred to as Ethernet+Cloud and 5G+Cloud), we introduce an additional baseline called 802.11ac+Cloud. This baseline replaces the Ethernet/5G I2I method with campus Wi-Fi.}

%% file: 07-2-Exp-Overall.tex
\vspace{-1em}
\subsection{Overall Performance}\label{subsec:exp_overall}
\begin{table*}[t!] 
%\scriptsize
\small
\caption{Application settings and distribution. Text in italics describes the settings for lite models.} 
\label{tab:outdoor_para}
\vspace{-1.5em}
\begin{tabular}{c|c|c|c|c|c|c} 
\hline
\hline
{\bf Application } & {\bf Source Node} & {\bf Model} & {\bf Sensor}& {\bf Execution} & {\bf \blue{Deadline (ms)}}& {\bf Accuracy} \\
\hline
Perception Sharing (P) & 1,3,7,9,11 & PointPillars & LiDAR & 128.0 /\textit{118.3}\,ms & \blue{250 / 300 / 350} & $85.5\%$ / $\textit{83.0\%}$\\
\hline
Traffic Monitoring (T) & 1,2,5,7,10 & YOLO & Virtual Camera & 37.2 /\textit{32.7}\,ms & \blue{300 / 350 / 400} & $87.1\%$ / $\textit{81.0\%}$\\
\hline
Jaywalk Warning (J) & 3,4,5 & YOLO & Thermal & 78.3 /\textit{43.3}\,ms  & \blue{250 / 300 / 350}  & $84.7\%$ / $\textit{82.0\%}$\\
\hline
\hline
\end{tabular}
\vspace{-1em}
\end{table*}

\begin{table*}[htb]
%\tiny
%\normalsize
\small
\centering
\caption{Task allocation results on smart lampposts.}
\label{tab:task-alloct}
\vspace{-1.5em}
\begin{tabular}{p{1.3cm}|p{1cm}|p{0.9cm}|p{1.0cm}|p{0.9cm}|p{1.0cm}|p{1.0cm}|p{1.0cm}|p{0.9cm}|p{0.9cm}|p{0.9cm}|p{0.9cm}|p{0.9cm}} 
\hline
\hline
& Node1 & Node2 & Node3 & Node4 & Node5 & Node6 & Node7 & Node8 & Node9 & Node10 & Node11 & Node12\\ 
\hline
Baseline & P1,T1 & / &P2,J1,J2 & J3,J4 & T2,T3,J5 & / & P3,P4,T4 & / & P5 & T5,T6 & P6,P7& /
\\ 
\hline
% \sysname & P1,T1 & J1,J2 & P2 & J4,T2 & J3,T3 & J5,T4 & P3 & P4 & P5 & T5,T6 & P7 & P6 \\ 
\sysname & P1,T1 & \blue{J1} & \blue{P2,J2} & \blue{J4} & \blue{J3,J5} & \blue{T2,T3,T4} & P3 & P4 & P5 & T5,T6 & P6 & P7 \\ 
\hline
\hline
\end{tabular}
\vspace{-1em}
\end{table*}

\myparagraph{Evaluation setup.}
We evaluate \sysname with varying combinations of applications across nodes of Cluster\,2.
\sysname nodes have varying task sets since they have different combinations of sensors. Table~\ref{tab:outdoor_para} presents the distribution of tasks on nodes.
The \textit{source node} is where the data required by the task is generated, which is set according to the sensors and the covered road condition (e.g., crossings).
\blue{\sysname aims to extend the field of view for AVs, enabling them to have a broader perception. 
The \sysname node shares application results with the vehicle when it enters the sensing range.
We set the \textit{deadline} for the delivery of results so that the vehicles can receive a considerable portion of the SRI's view. As an example, we now derive the deadline setting for a vehicle entering the sensing range of a \sysname node (e.g., $50\,m$ sensing range) at a speed of $70\,km/h$. 
We assume that the \sysname node must share at least $90\%$ of the sensing range (i.e., 45 out of the 50-meter range). That is, the data from \sysname must be received by the vehicle when it travels no more than 5 meters into the sensing range of \sysname node, which is equivalent to a time duration of $\frac{5\,m}{70\,km/h} = 250\,ms$.
In our experiments, the sensors on \sysname typically have a sensing range of $50 \sim 70\,m$. 
To ensure that vehicles traveling at speeds between $50 \sim 70\,km/h$ share at least $90\%$ of \sysname node's sensing range, we set application deadlines in the range of $250 \sim 500\,ms$.
Specifically, we set three different levels of application deadlines as outlined in Table~\ref{tab:outdoor_para}, with the most urgent deadline set at $250\,ms$ for perception sharing and jaywalk warning, and $300\,ms$ for traffic monitoring, as the former applications typically require more urgent attention.
\new{We note that \sysname essentially functions as an on-the-air sensor for vehicles to provide additional perception beyond the short range around the vehicles. Since most autonomous vehicles only rely on their own sensors to perceive the short-range surroundings, \sysname does not affect the safe-critical short-range path planning.}
}

\begin{figure}[t!]
	\begin{minipage}[b][][b]{\columnwidth}
		\centering
		\includegraphics[width=.9\columnwidth]{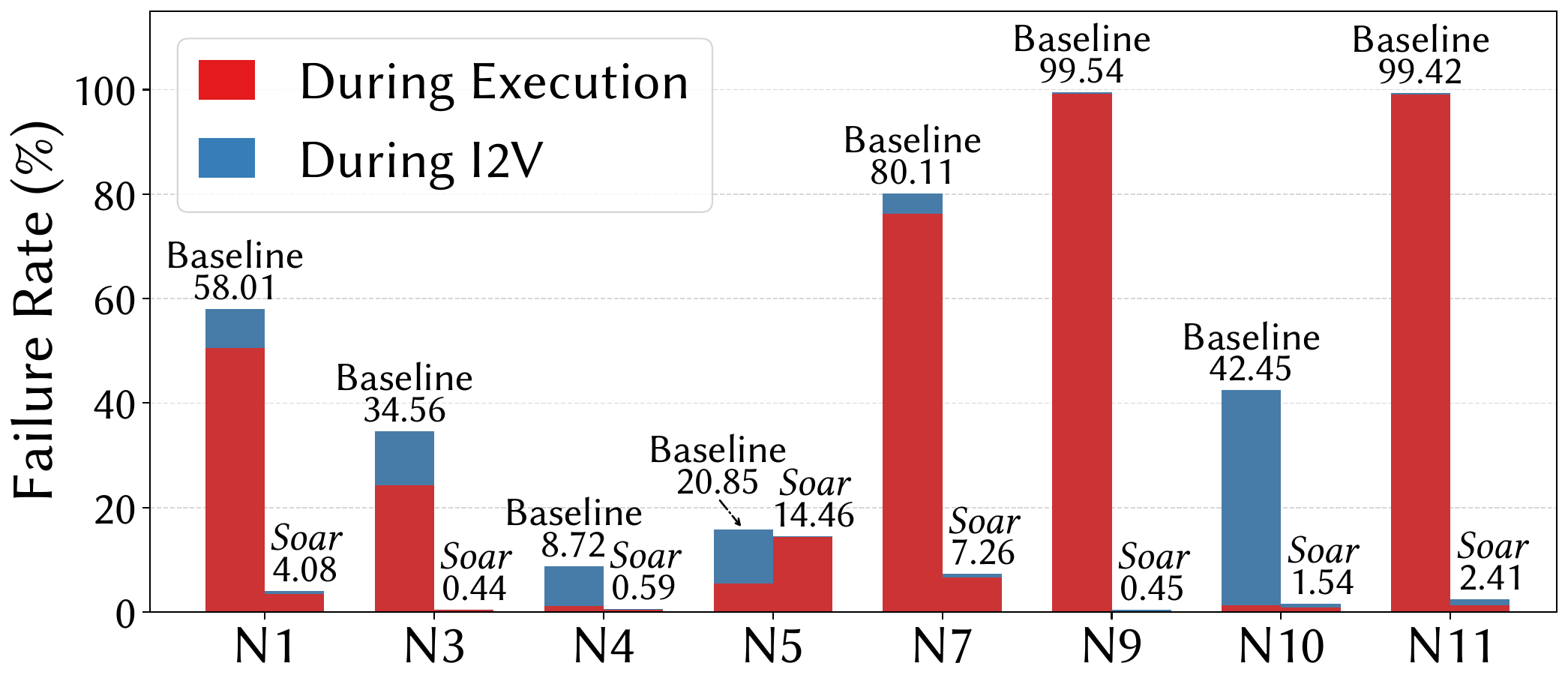}\\[-1ex]
	\end{minipage}\vspace{-1.5em}
	\caption{\blue{Overall performance of \sysname system.}}
    \label{fig:overall-perf}
    \vspace{-2em}
\end{figure}

\myparagraph{Overall task performance.} 
We evaluate the overall application performance by comparing \sysname with the overall baseline described in \S~\ref{subsec:methodology}. \blue{We provide a comprehensive analysis under the most urgent deadline setting, and summarize the performance improvement achieved for the other two deadline settings.} Table~\ref{tab:task-alloct} shows the allocation strategy generated by \sysname. We observe that \sysname successfully offloads the DL tasks from the heavily loaded node to the idle or lightly loaded node. For example, a data-intensive PointPillars-based DL task migrated from Node\,11 to Node\,12. 
Fig.~\ref{fig:overall-perf} shows the failure rate for each \sysname node. We omit several \sysname nodes in Fig.~\ref{fig:overall-perf}  as we calculate the failure rate based on tasks' source node, and no task is sourced from these omitted nodes. 
\blue{We observe that \sysname can achieve a failure rate reduction at $41.19\%$ on average before transmitting the application results. After transmitting the application results, \sysname can maintain a failure rate below $14.46\%$ ($3.90\%$ on average), while the baseline incurs a failure rate up of $99.54\%$. In summary, \sysname reduces the failure rate significantly ($50.93\%$ on average) compared with the baseline.}
\blue{For the other two deadline settings, \sysname demonstrates an average reduction in the failure rate of $28.39\%$ and 
$24.40\%$, respectively.}
The results show that \sysname consistently maintains a reliable application performance for AVs among multiple \sysname nodes.

% \vspace{-1em}
\begin{figure}[t!]
\centering
    \begin{subfigure}[b]{0.21\linewidth}
        \includegraphics[width=\textwidth]{{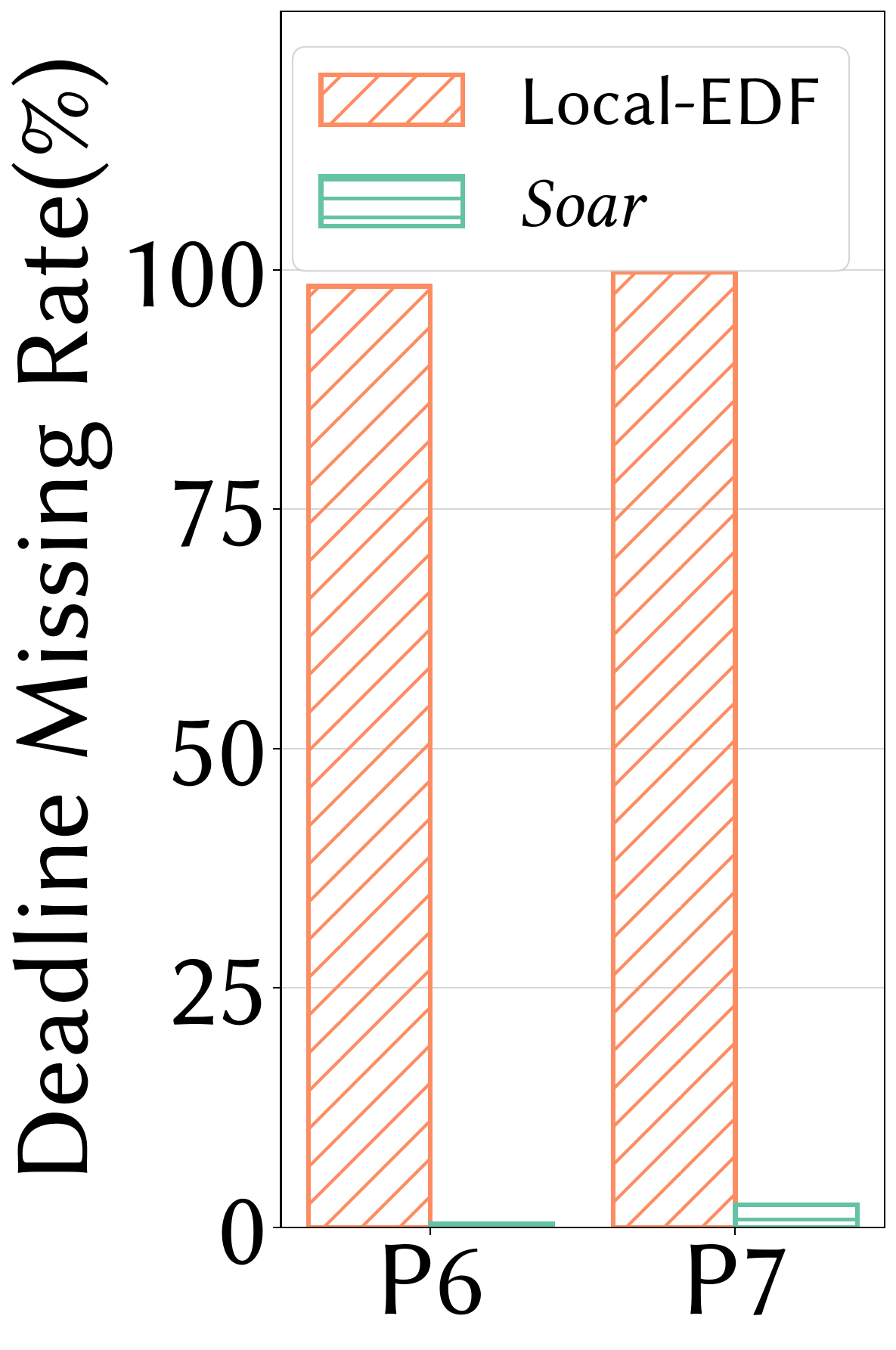}}
        \vspace{-2em}
        \caption{\blue{Node\,11}}
        \label{fig:node7_miss}
    \end{subfigure}
    \hfill
    \begin{subfigure}[b]{0.21\linewidth}
        \includegraphics[width=\textwidth]{{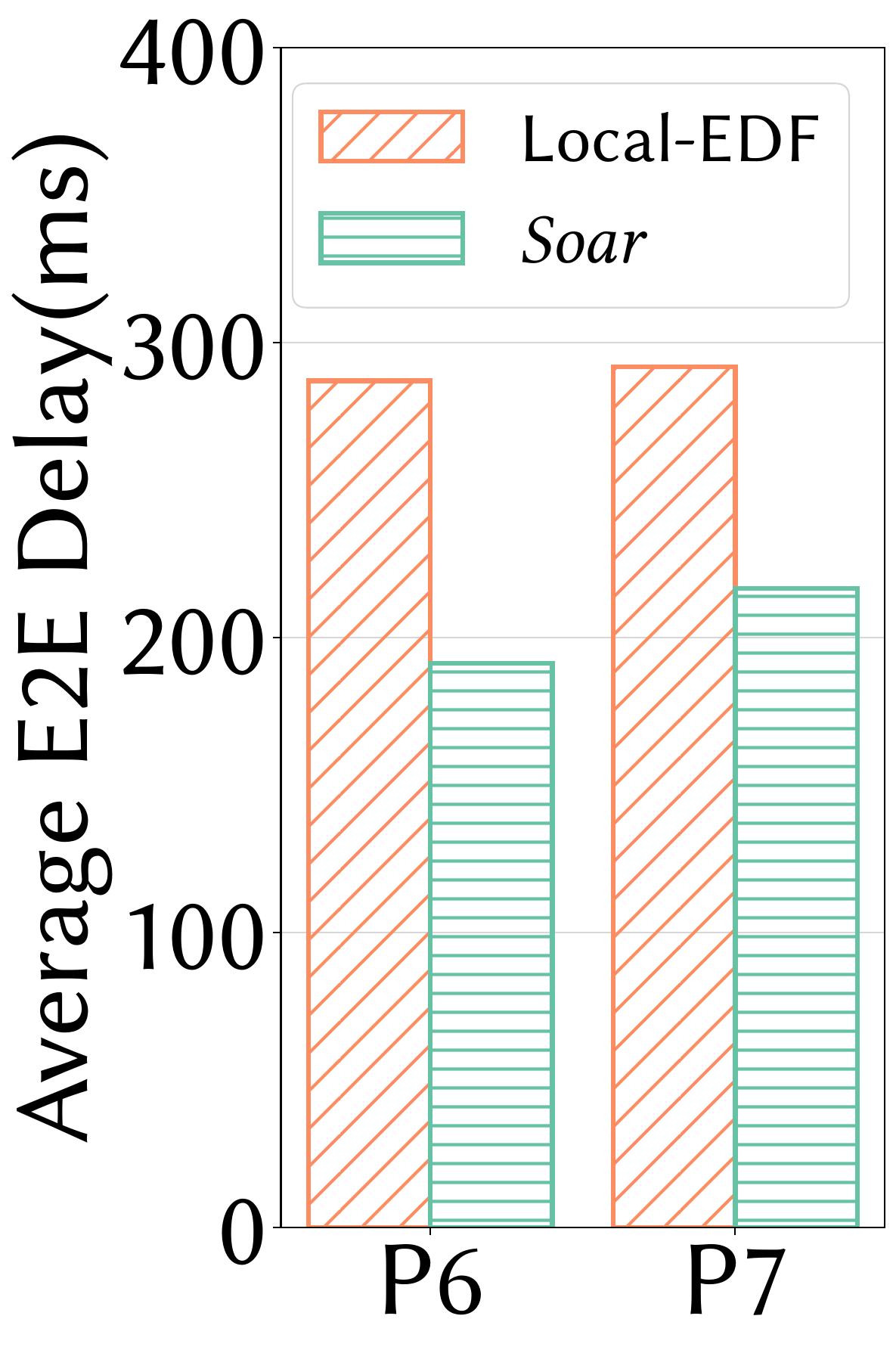}}
        \vspace{-2em}
        \caption{\blue{Node\,11}}
        \label{fig:node7_avg}
    \end{subfigure}
    \hfill
    \begin{subfigure}[b]{0.21\linewidth}
        \includegraphics[width=\textwidth]{{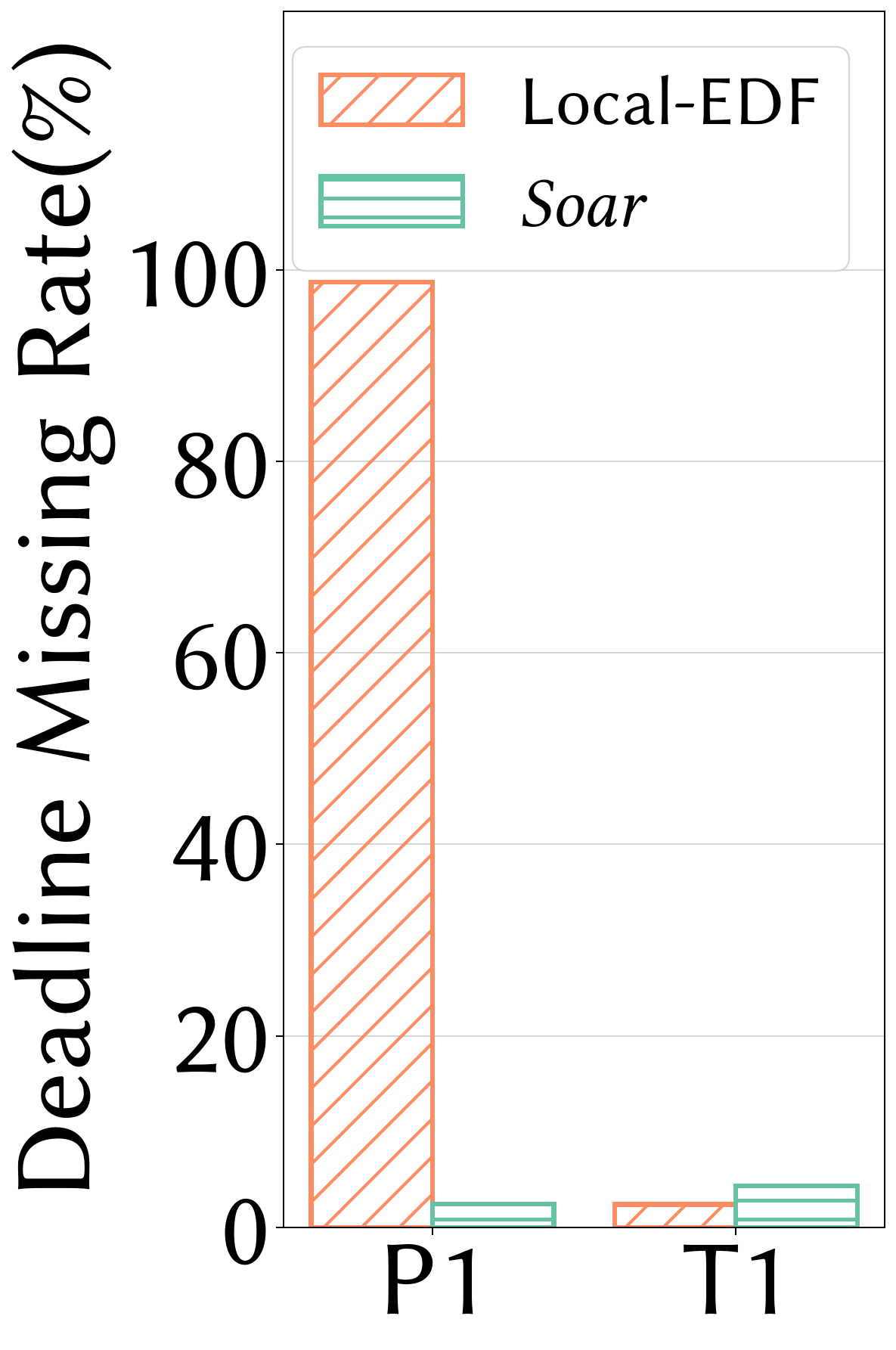}}
        \vspace{-2em}
        \caption{\blue{Node\,1}}
        \label{fig:node1_miss}
    \end{subfigure}
    \hfill
    \begin{subfigure}[b]{0.21\linewidth}
        \includegraphics[width=\textwidth]{{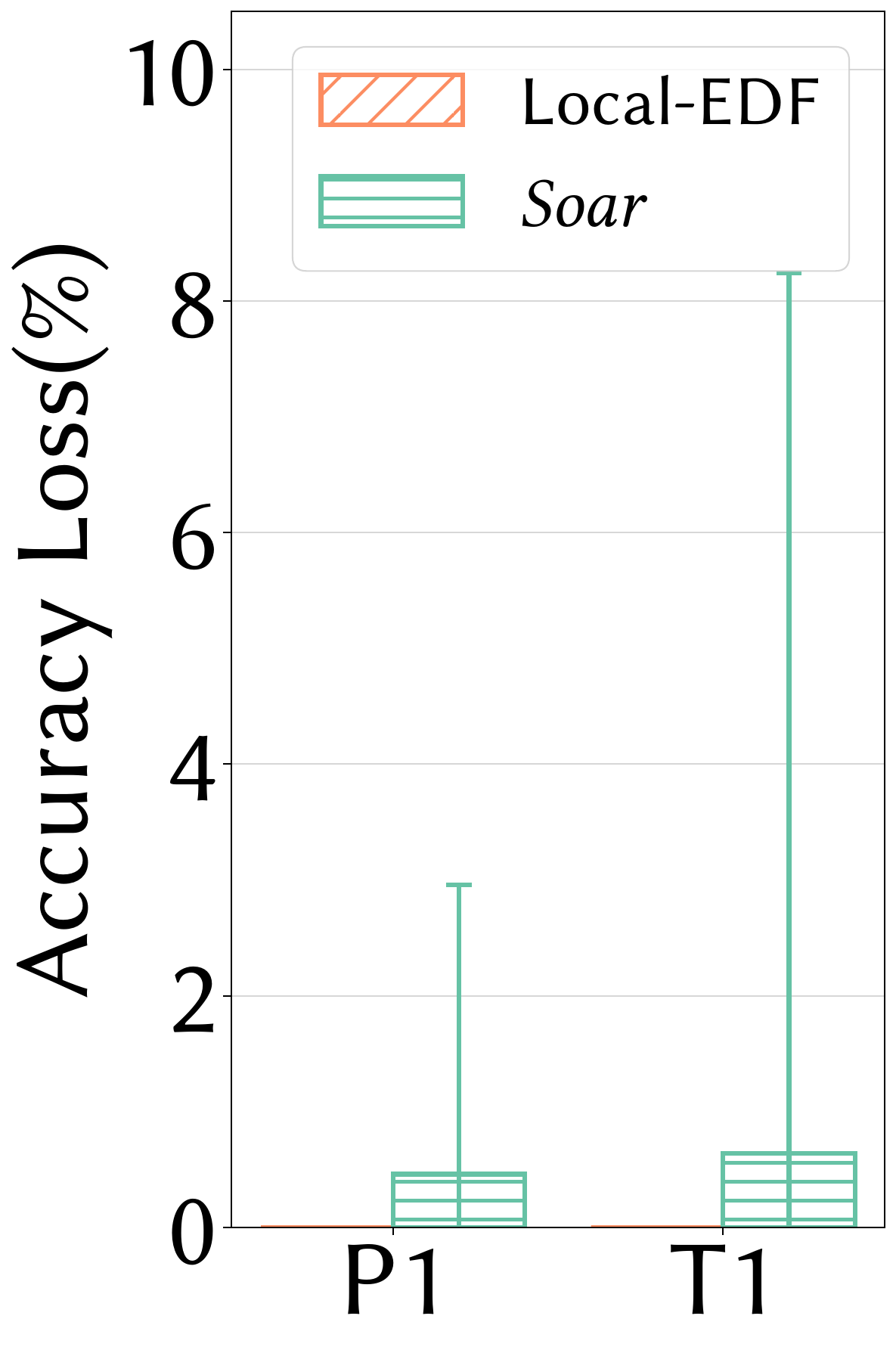}}
        \vspace{-2em}
        \caption{\blue{Node\,1}}
        \label{fig:node1_acc}
    \end{subfigure}
    \vspace{-1em}
    \caption{\blue{Performance of task management.}}
    \label{fig:overall_computing_res}
    \vspace{-2em}
\end{figure}

We further analyze the advantages of task management in \sysname by focusing on the real-time performance of each task on a single node.
Fig.~\ref{fig:overall_computing_res} shows the deadline missing rate and average E2E delay of each task sourced from Node\,11. \sysname reduces deadline missing rate efficiently by \blue{$97.67\%$}. The average E2E delay is decreased by \blue{$85.43\,ms$} since our task management has successfully allocated \blue{the task $P7$ from heavily loaded Node\,11 to the idle Node\,12.} 
We also show the deadline missing rate and accuracy loss on Node\,1. The results show that \sysname can reduce the deadline missing rate by \blue{$47.16\%$} on average with all tasks executed locally. 
This is because our opportunistic DL task scheduler efficiently chooses the lite model for execution when the remaining time for the current and next inference is about to run out, which only leads to \blue{$0.55\%$} accuracy loss on average.

\begin{figure}[!t]
		\centering
		\includegraphics[width=.9\columnwidth]{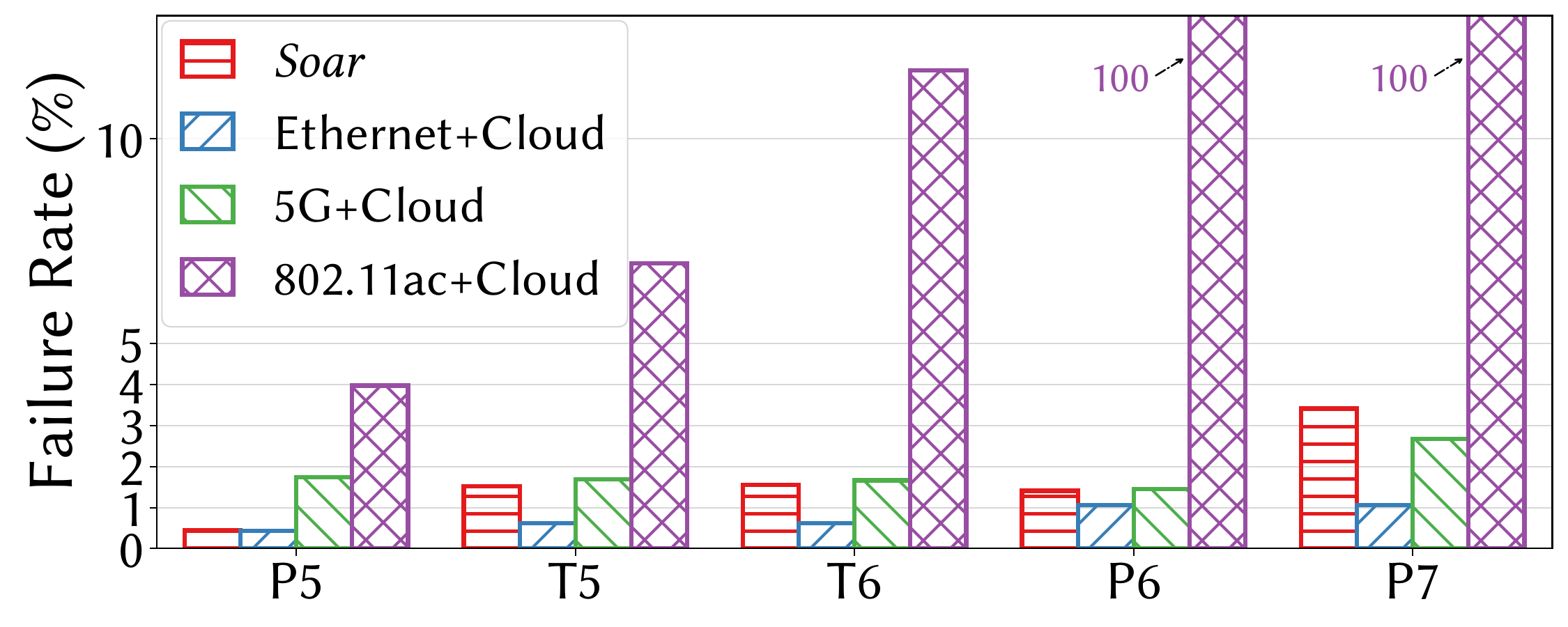}
    \vspace{-1.5em}
	\caption{\blue{Comparison with cloud baselines.}}
        \vspace{-2em}  
	\label{fig:res_cloud}
\end{figure}

\myparagraph{\blue{Performance comparison with cloud baselines.}}
\blue{We evaluate the performance between \sysname and cloud baselines (see \S~\ref{subsec:methodology}) using different network configurations.
Since our outdoor testbed supports limited communication settings, we set up an indoor testbed that supports the simulation of diverse network conditions.
Specifically, three TX2 edge nodes transmit their sensor data to an NVIDIA GeForce RTX 2080 Ti server for processing through a wired Ethernet connection.
\new{We implement the baselines for cloud-based implementations by applying a trace collected from a public local AWS~\cite{AWS} server using \textit{CloudPing}~\cite{cloudping}, in which the result exhibits a $15\,ms$ round-trip time (RTT) between \sysname nodes and the cloud on average.} 
To emulate the 5G+Cloud and 802.11ac+Cloud communication, we capture real-world traces using \textit{SpeedTest}~\cite{speedtest} at each \sysname node of our outdoor testbed. These traces contain bandwidths and dynamics of 5G and campus Wi-Fi.
We replicate the settings used for Node\,9, 10, 11 from Fig.~\ref{fig:overall-perf} and show the failure rate of each task in Fig.~\ref{fig:res_cloud}.  
The results indicate that Ethernet+Cloud outperforms \sysname in terms of performance but comes with higher deployment costs, as discussed in \S~\ref{subsec:design-choice}. 
However, \sysname demonstrates an average reduction in failure rate of $0.17\%$ and $42.85\%$ compared to 5G+Cloud and 802.11ac+Cloud, respectively. 
802.11ac+Cloud performs poorly because many tasks failed to meet the deadlines due to significant delays caused by the transmission of large volumes of raw sensor data.
Although 5G is capable of achieving high-bandwidth communication, our traces indicate that its uplink performance was unsatisfactory, only reaching $80\,Mbps$ on average.
Furthermore, we observe that the network may suffer significant bandwidth degradation (i.e., $40\,Mbps$) due to various blockages. Such limited bandwidth leads to significant delays, resulting in a high failure rate.
\new{Besides performance gain, \sysname incurs lower operation costs (c.f., Table~\ref{tab:cost}). Although cloud-based alternatives reduce the maintenance cost of computation units, the costs associated with repairing or replacing faulty sensors and communication devices are inevitable. In contrast, \sysname avoids costly 5G operations, base station installation, and cable maintenance, making it a viable and practical solution for long-term deployment.}}

\myparagraph{Performance of applications on the vehicle.}
We evaluate the \sysname can support infrastructure-assisted autonomous driving applications on the vehicle side. Table~\ref{tab:vi-eye} and Table~\ref{tab:vips} present the results of point cloud registration and LiDAR perception extension, respectively. The ``offline'' in the tables denotes the offline execution performance of our implemented applications using the original point clouds from the vehicle and infrastructure sides. \new{Table 4 uses ``success rate" and ``density benefit" to present the performance of point-cloud registration and beneficial sensor data for vehicles, respectively. Table~\ref{tab:vips} leverages ``fps" to show the frame rate of the perception fusion. We note that these metrics are widely adopted in various work~\cite{shi2022vips, he2021vi}.}

The point cloud registration requires \sysname to transmit a large volume of raw point clouds to the vehicle, so its performance indicates the I2V communication performance. Results in Table~\ref{tab:vi-eye} indicate that using point clouds transmitted by \sysname with the error correction code (ECC), the vehicle can achieve a registration performance and point cloud density benefit similar to using original point clouds. This is because passive data transmission can suffer from severe packet loss without ECC. 
For the LiDAR perception extension, it requires only negligible results transmission while introducing a data-intensive task (i.e., 3D object detection with point cloud) to \sysname. Table~\ref{tab:vips} shows that \sysname achieves consistent perception fusion accuracy and the ratio of perception extension among different hardware platforms compared with the baseline.
However, the baseline can only achieve 1.73 fps which incurs significant perception fusion errors (over 1\,m) and benefit loss.
The reason that \sysname can achieve an acceptable fps \new{(i.e., $> 3$ fps) required by the application~\cite{shi2022vips}} is the optimal task execution by our task management framework. 

\begin{table}[!t]
% \begin{minipage}[t]{.8\columnwidth}
    \centering
    \caption{Point cloud registration on \sysname.}
    \label{tab:vi-eye}
    \vspace{-1em}
    \resizebox{.8\columnwidth}{!}{%
    \begin{tabular}{c|c|c|c}
    \hline
    \hline
    Methods                         & Offline & \sysname & \sysname w/o ECC \\ \hline
    Registration Error (m)          & 0.10                  & 0.16     & 0.39                \\ \hline
    Success Rate (\%)               & 85.32                 & 77.31    & 52.43               \\ \hline
    Density Benefit ($pts/m^2$)     & 46                    & 39       & 10                  \\ \hline
    \hline
    \end{tabular}%
    }
    \vspace{-1.5em}
% \end{minipage}
\end{table}
% \vspace{em}

\begin{table}[!t]
\hspace{.5em}
% \begin{minipage}[t]{.9\columnwidth}
    \centering
    \caption{LiDAR perception extension on \sysname.}
    \label{tab:vips}
    \vspace{-1em}
    \resizebox{.9\columnwidth}{!}{%
    \begin{tabular}{c|c|c|c}
    \hline
    \hline
    Methods            & Offline (Laptop) & \sysname (TX2) & w/o \sysname (TX2)  \\ \hline
fps                & 10               & 3.67           & 1.73                 \\ \hline
Fusion Error (m)   & 0.46             & 0.73           & 1.09                \\ \hline
Benefit Ratio (\%) & 42.1             & 37.9           & 15.4               \\ \hline
    \hline
    \end{tabular}%
    }
\vspace{-1.5em}
% \end{minipage}
\end{table}
% \vspace{-1em}

%% file: 07-3-Exp-comm.tex
\begin{figure*}[!t]
    \centering
    \begin{minipage}[b][][b]{.65\columnwidth}
		\centering
		\includegraphics[width=\columnwidth]{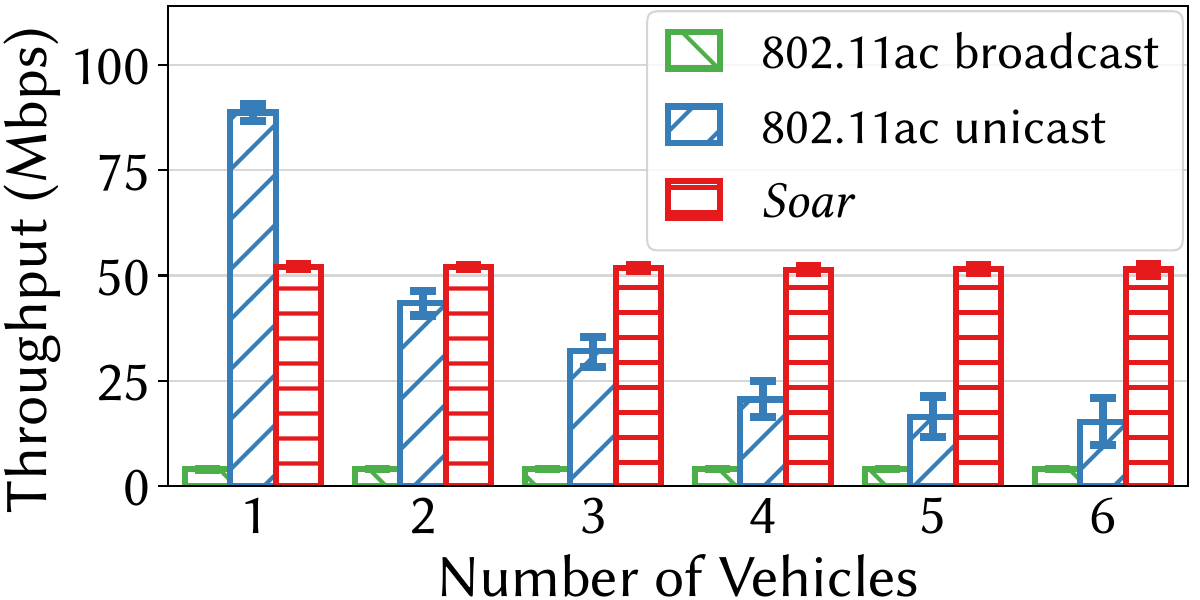}
        \vspace{-2.5em}
	    \captionof{figure}{Throughput vs the number of vehicles.}
        \vspace{-1.5em}
	\label{fig:res_multicar}
    \end{minipage}
    \hspace{.5em}
    \begin{minipage}[b][][b]{.65\columnwidth}
		\centering
		\includegraphics[width=\columnwidth]{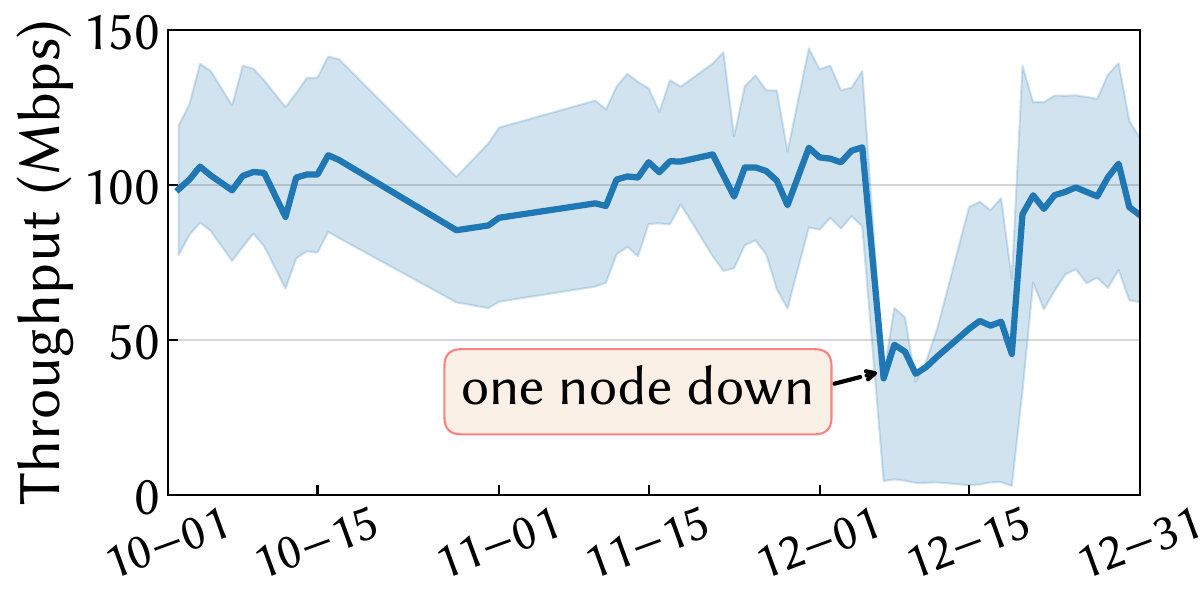}
        \vspace{-2.5em}
	    \captionof{figure}{\blue{3-month throughput evaluation of \sysname in 2022.}}
        \vspace{-1.5em}
	\label{fig:long_term}
    \end{minipage}
    \hspace{.5em}
    \begin{minipage}[b][][b]{.65\columnwidth}
		\centering
		\includegraphics[width=\columnwidth]{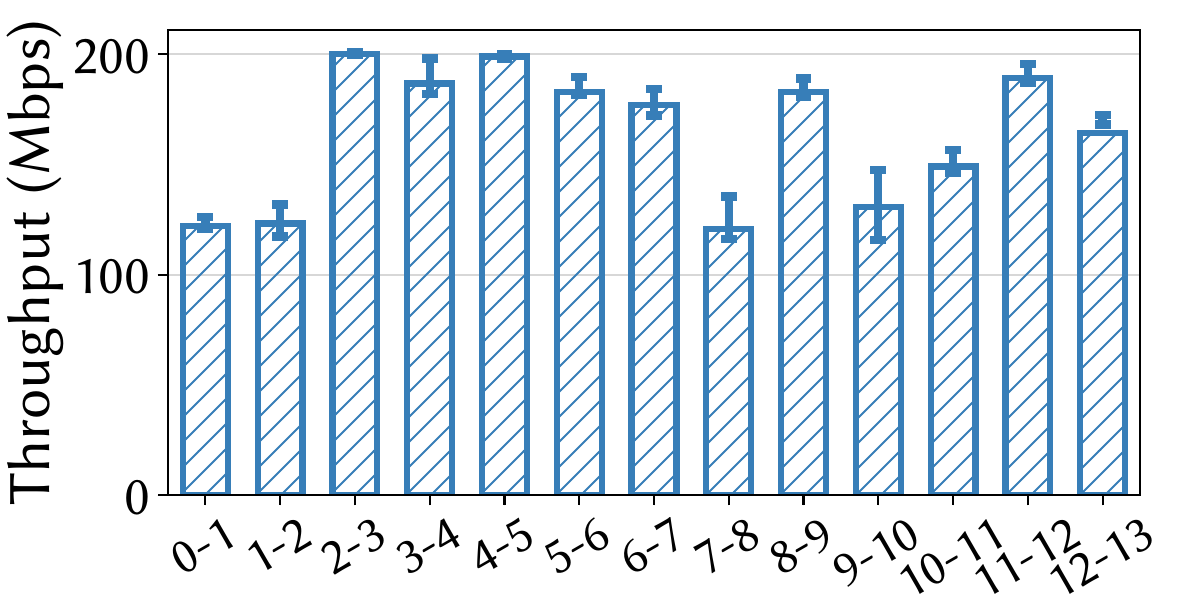}
        \vspace{-2.5em}
	    \captionof{figure}{\blue{Throughput of each link of \sysname on Oct. 2022.}}
        \vspace{-1.5em}
	\label{fig:diff_node}
    \end{minipage}
    \hspace{.5em}
\end{figure*}

\begin{figure}[htb]
    \begin{minipage}[t]{.58\columnwidth}
	\centering
        \includegraphics[width=\columnwidth]{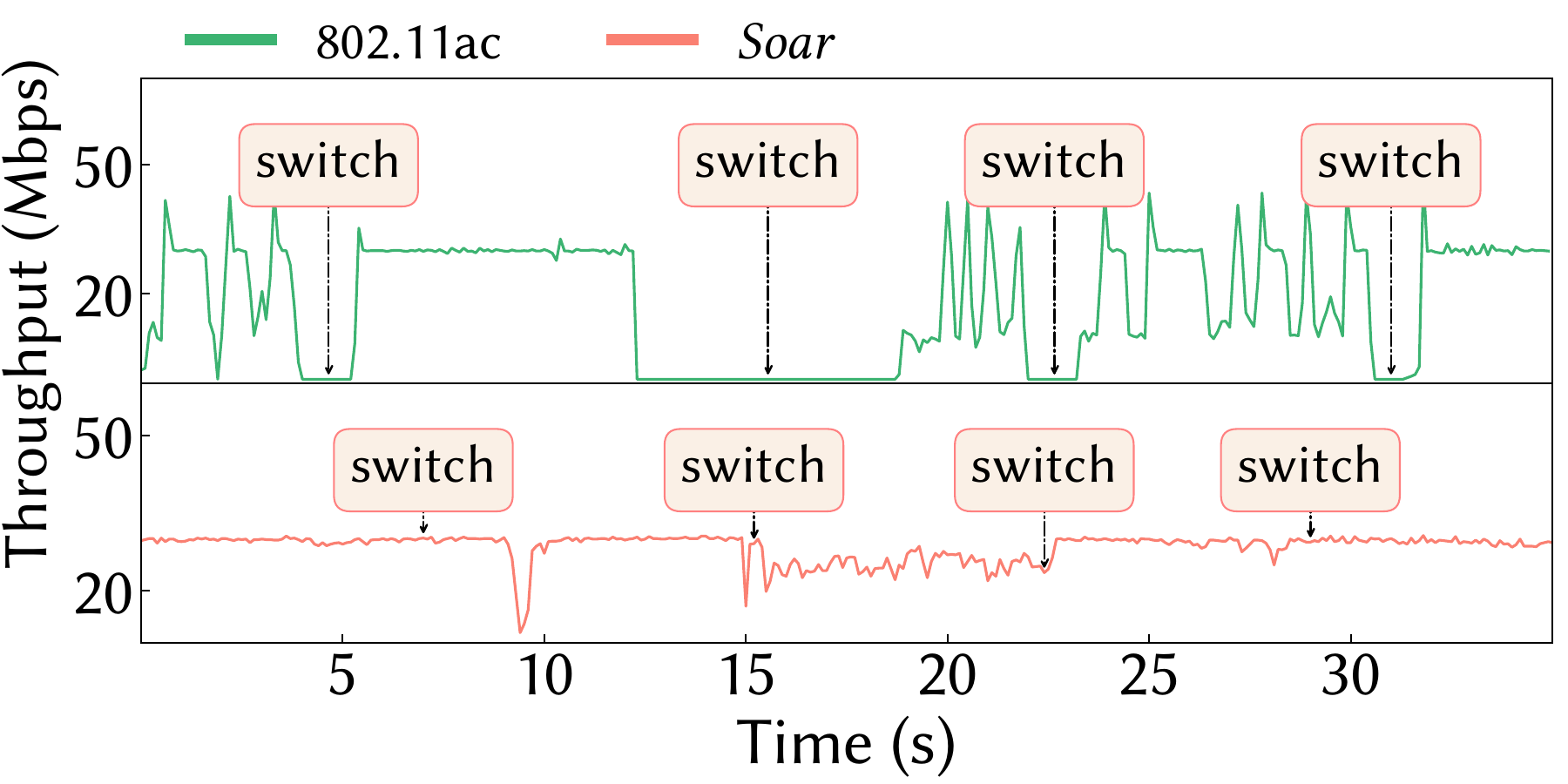}
        \vspace{-2.5em}
	\caption{An example of switching over \sysname nodes.}
	\label{fig:res_switch}
    \end{minipage}
    \hspace{0.04cm}
    \begin{minipage}[t]{.38\columnwidth}
	\centering
        \includegraphics[width=\columnwidth]{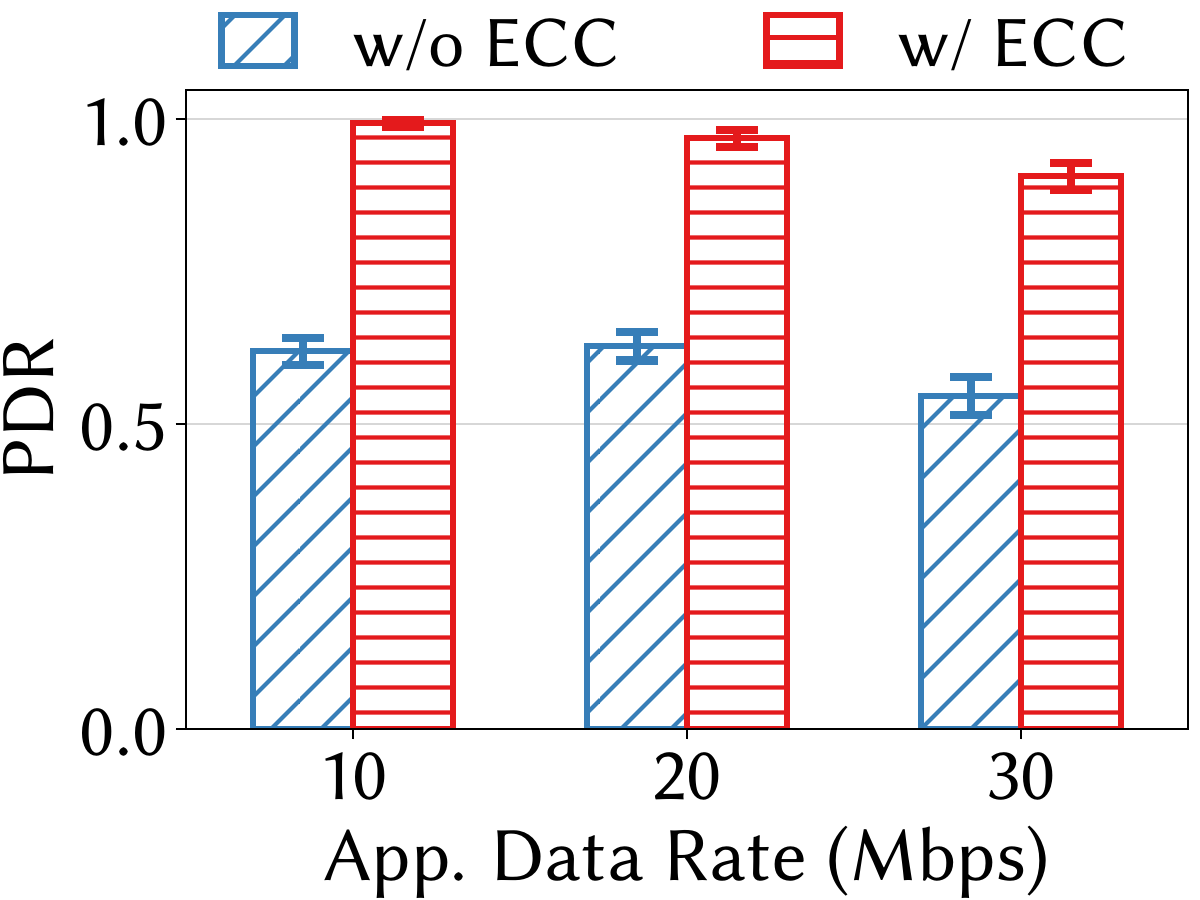}
        \vspace{-2.5em}
	\caption{Performance of ECC.}
	\label{fig:res_coding}
    \end{minipage}
    \vspace{-2em}
\end{figure}
 
% \vspace{-1em}
\subsection{Communication Benchmark}
\label{subsec:exp_comm}
\myparagraph{Impact of the number of vehicles.} We first investigate the impact of the number of vehicles on I2V communication. We employ a vehicle with different numbers of pseudo vehicles. The speed of vehicles is around $10\sim20$\,km/h. Fig.~\ref{fig:res_multicar} shows the throughput of \sysname and two baselines with different numbers of vehicles. \sysname and 802.11ac broadcast exhibit negligible change, while the throughput of 802.11ac unicast plummeted with more vehicles.
Although 802.11ac broadcast is resistant to the vehicle number, its low throughput can not afford the large volume of data required by autonomous driving applications.
On the other hand, 802.11ac unicast presents terrible scalability. 
\new{In contrast, \sysname demonstrates consistent throughputs exceeding $50$\,Mbps regardless of the number of vehicles. To be specific, when there are six vehicles, \sysname achieves $3\times$ the throughput of 802.11ac unicast.}

\myparagraph{Impact of switching overhead.}
We further explore the impact of switching overhead on the performance of our I2V communication. We utilize five consecutive \sysname nodes to transfer raw LiDAR data at $30$\,Mbps. One vehicle drives through the nodes at different speeds. We choose 802.11ac as the baseline. 
Fig.~\ref{fig:res_switch} shows a data trace of channel switching when the vehicle moves at $20$\,km/h. The upper and lower lines are the throughput of 802.11ac and \sysname, respectively. 
802.11ac exhibits significant switching overhead because of its prolonged reassociation.
In addition, our measurement in Fig.~\ref{fig:switch_stats} shows that 802.11ac suffers from non-trivial switching overhead with higher vehicle speed. 
Such a high overhead prevents 802.11ac from being used in our I2V communication.
However, \sysname only requires adjusting the sniff channel with negligible performance degradation, which enables \sysname to provide seamless data transmission to vehicles.

\myparagraph{Error correction code.}
We evaluate the performance of the error correction code used in our passive downlink I2V broadcast. 
Fig.~\ref{fig:res_coding} shows the PDR results under various application data rates.
When the data rate is low (e.g., $10$\,Mbps), \sysname with error correction code can achieve less than $1\%$ packet loss. With a large amount of data (e.g., $30$\,Mbps), we can still maintain over $90\%$ of PDR. Instead, without error correction codes, the PDRs are only around $60\%$.
The results show that \sysname with error correction codes can achieve high-goodput and reliable transmission in different applications.

%% file: 07-4-Exp-sys-eva.tex
\vspace{-2.5em}
\subsection{System-level Performance}
\blue{We present long-term system-level performance of \sysname in this section.
First, we evaluate the long-term communication performance of our deployment by collecting throughput from each node to the sink node between Oct. 2022 and Dec. 2022.
Fig.~\ref{fig:long_term} shows that the average throughput consistently achieved $100\,Mbps$, except a notable performance degradation occurred on 7 Dec. 2022 due to the failure of one node.
However, with our semi-fixed routing strategy (see \S~\ref{subsec:I2I}) nearby \sysname nodes performed link recovery to establish new links, which allows \sysname to still maintain around $50\,Mbps$ during the transition.
The error bar in Fig.~\ref{fig:long_term} shows a percentile interval from $25\%$ to $75\%$, which illustrates a considerable level of link instability (e.g., over $50\,Mbps$).}
\blue{To investigate the performance of each link of \sysname nodes, we analyze the per-link throughput of \sysname Cluster 2 in Oct. 2022.
Fig.~\ref{fig:diff_node} shows that the throughput of links varies substantially, which is caused by the differences and dynamics in the physical environment of each node.
Nevertheless, \sysname can achieve over $100\,Mbps$ throughput for all links.
}

\blue{
It is well known that the CPU/GPU performance can degrade considerably at high temperatures~\cite{ThermalThrottling}. We further tested the computing performance of the CPU and GPU on \sysname over a period of three months, with an average temperature of $20 \sim 29 ^{\circ} C$.
We use \textit{sysbench}~\cite{Sysbench} and \textit{mixbench}~\cite{konstantinidis2017quantitative} to evaluate the CPU and GPU performance, respectively.
The experimental results reveal that the average execution time for a \textit{sysbench} event on CPU ranges from $90.88\,ms$ to $90.91\,ms$, indicating no significant performance degradation.
The GPU performance is measured to be $20.79\,GFLOPS$ on average, with a standard deviation of $0.005$. 
Overall, the results demonstrate no significant performance fluctuations for the CPU/GPU of our platforms over several months despite varying temperatures and prolonged use.}

%% file: 08-Discussion.tex
\section{Discussion}
\label{sec:discussion}

\myparagraph{\new{Cost-effectiveness.}}
\new{\sysname provides a reference design of SRI using inexpensive off-the-shelf computing and sensing components based on practical power and budget constraints of operational lampposts. \sysname's modular hardware design and data management framework enable cost-effective part replacement and plug-and-play functionality with powerful processors and multi-modal sensors. This design supports different components (e.g., communication, computing, etc.) to be upgraded in a cost-effective manner. In addition, \sysname nodes can be deployed with diverse densities to enable a trade-off between the assistive services for vehicles and the deployment cost. For example, \sysname nodes can be strategically deployed densely where they are most needed, such as busy intersections and crosswalks.}

\myparagraph{\new{Scalability.}}
 \new{\sysname's cluster-based networking and task management architecture facilitates large-scale deployments without compromising on application performance. \sysname efficiently broadcasts data in universal formats, such as raw sensor samples or bounding boxes, which seamlessly supports a majority of current autonomous driving applications. This paradigm is inherently scalable, as it simplifies the data dissemination process and ensures that \sysname can accommodate an expanding network of vehicles.}

\myparagraph{\new{Security Issues.}}
\new{In this work, we assume that \sysname functions under secure settings and the data transmission between infrastructure and vehicles is trusted. In real-world operations, various security measures can be integrated to enhance the security of \sysname, such as encryption of sensor data~\cite{guerrero2020blockchain}, safeguarding model inference with access controls~\cite{he2020attacking}, and authentication protocols for communication~\cite{bagga2020authentication}.}

%% file: 09-Conclusion.tex
\section{Conclusion}
\label{sec:conclustion}
This paper presents the design and deployment of \sysname, the first end-to-end SRI system specifically designed for supporting AVs. \sysname consists of carefully designed components for data and DL task management, I2I and I2V communication, and an integrated hardware platform, which addresses a multitude set of system and physical challenges, allows to leverage the existing operational traffic infrastructure, and hence lowers the barrier of adoption. Based on a real-world deployment of 18 \sysname nodes on existing lampposts on campus, our evaluation shows that \sysname can support a diverse set of autonomous driving applications, and achieve desirable real-time performance and high communication reliability. 

Our experience offers key insights into the development and deployment of next-generation SRI and autonomous driving systems. \sysname demonstrates a highly efficient yet practical instance in a fairly large design space for SRI. In the future, we will integrate emerging open communication systems based on advanced V2X technologies. In particular, an open question is how to implement high-bandwidth I2I and I2V communication in a cost-effective manner using 5G and emerging 6G cellular technologies. Lastly, we will investigate new cyber-physical security mechanisms that can protect AVs, which become increasingly critical with the prominence of smart roadside infrastructure.